\documentclass[a4paper,11pt]{article}
\pdfoutput=1
\usepackage{jinstpub}

\usepackage{natbib}
\usepackage{siunitx}
\usepackage{pgf}
\makeatletter
\@namedef{ver@amsmath.sty}{}
\makeatother
\usepackage{amstext}
\usepackage{chemformula}
\DeclareChemSymbol{e}{e}

\usepackage{hepnames}
\usepackage{url}
\usepackage{upgreek}
\usepackage{textalpha}

\newcommand{\upd}{\mathrm{d}}

\title{Monte Carlo simulations of the electron - gas interactions in the KATRIN experiment}

\author[a]{J.~Kellerer}
\author[b]{and F.~Spanier}

\affiliation[a]{Institut f\"ur Astroteilchenphysik, KIT, Hermann-von-Helmholtz-Platz 1,
76344 Eggenstein-Leopoldshafen, Germany}
\affiliation[b]{Institut f\"ur Theoretische Astrophysik, Universit\"at Heidelberg, Albert-Ueberle-Str. 2 und Philosophenweg 12,
69120 Heidelberg, Germany}

\emailAdd{felix@fspanier.de}

\abstract{
At the KATRIN experiment, the electron antineutrino mass is inferred from the shape of the $\upbeta$-decay spectrum of tritium. Important systematic effects in the Windowless Gaseous Tritium Source (WGTS) of the experiment include the energy loss by electron scattering, and the extended starting potential. In the WGTS, primary high-energy electrons from $\upbeta$-decay produce an extended secondary spectrum of electrons through various atomic and molecular processes including ionization, recombination, cluster formation and scattering. In addition to providing data essential to the simulation of energy loss processes, the electron spectrum also provides information important in the simulation of plasma processes. These simulations will then provide an insight on the starting potential.\\
Here, a Monte Carlo approach is used to model the electron spectrum in the source for a given magnetic and electric field configuration. The spectrum is evaluated at different positions within the WGTS, which allows for a direct analysis of the spectrum close to the rear wall and detector end of the experiment. Alongside electrons, also ions are tracked by the simulation, resulting in a full description of the currents in the source.
}


\begin{document}
\maketitle
\flushbottom

\section{Introduction}

The \textbf{KA}rlsruhe \textbf{TRI}tium \textbf{N}eutrino (KATRIN) experiment is designed to measure the electron antineutrino mass to unprecedented sensitivity down to \SI{0.2}{eV} \citep{KATRIN2005}. At the experiment, the mass is inferred from the shape of the $\upbeta$-decay spectrum of tritium. The contribution of the mass on the shape is most dominant in the endpoint region. Thus, the KATRIN experiment is designed to provide a high energy resolution, as well as a high event rate in this region of interest. The energy resolution is reached through a spectrometer with an energy resolution of \SI{2.8}{eV} at \SI{18.6}{keV} kinetic energy of the electron \citep{Aker.1132021}. The measurement of the spectrum is performed down to \SI{40}{eV} below the endpoint \citep{Aker.1132021}. This part of the spectrum is well described by the convolution of the response function of the experiment and the theoretical beta spectrum, see \citep{Aker.1132021} for more information. This paper however focuses on the evaluation of charged particle properties far below the endpoint, in the high luminous tritium source, called the Windowless Gaseous Tritium Source (WGTS). Gaseous tritium is circulating at an ambient temperature of \SI{80}{\kelvin}, see section~\ref{sec:kat:over}. A strong magnetic field of \SI{2.5}{\tesla} is directed along the WGTS. The source tube of the WGTS is closed off at one side by the so-called rear wall, a gold-plated disc. At the other side, the source tube of the WGTS is continued in the differential pumping section (DPS), which houses additional pumps for tritium removal and electrodes for ion flux reduction.

When tritium decays, the resulting primary electrons follow the magnetic field lines and are either terminated directly at the rear wall or reach the DPS section of the experiment. All electrons except for those with high energy are repelled here at the dipole electrodes, which are at negative potential ($U_\text{DP,min} = \SI{-85}{V}$ \citep{Aker.352021}). The high energy component of the electrons reaches the spectrometer. However, only a small portion of these electrons (relative factor of approximately \num{E-10} \citep{Aker.352021,Aker:2019aa}) reach the detector, and contribute to the neutrino mass measurement. The majority is reflected back to the source. So in total, most of the electrons will eventually be terminated at the rear wall.

While the description of the electron movement seems straightforward, the tritium gas is also a target for the primary electrons from tritium beta decay, producing secondary electrons and altering the primary electron spectrum \citep{schimpf.2021}. This leads to a further complication; previous studies \citep{Nastoyashchii:2005aa,Kuckert.2016} have already shown that the density of electrons is sufficient to assume a plasma inside the WGTS. The plasma potential changes the relative energy of primary electrons to the detector. A precise knowledge of the electron spectrum and in turn also the plasma itself is necessary to understand the signal at the detector, see section~\ref{sec:kat:plasma}.

Studies of the total electron spectrum \citep{Nastoyashchii:2005aa} in the source exist for more than a decade, but have proven to be not sufficient to answer some questions: the electron spectrum at arbitrary positions (which is required for the simulation of kinetic plasmas) and the current of all charged particles (not only electrons) to the various walls of the experiment. Only some particle currents can be measured by the experiment, namely the total current towards the rear wall and the ion current towards the DPS \citep{Aker.352021}. Thus, the non-measurable currents must be determined through simulation.

The basic simulation concept of using a Monte Carlo approach has proven to be successful by Nastoyashchii et al. \citep{Nastoyashchii:2005aa}, but since then more precise interaction cross sections are known. Due to recombination processes and undetermined ion currents, which influence the development of the plasma, it is self-evident that the simulation by Nastoyashchii et al. of electrons alone does not provide a self-consistent picture of the processes in the WGTS.

In this paper, we will outline which physical processes are implemented in the newly developed code named KARL (and which cross sections are used to model these processes) and how this is realized in the numerical experiment. After a description of validation methods, the code is then used to provide electron spectra, particle densities and particle currents at multiple positions inside the source at current measurement conditions.

The code base for KARL was initiated in the Masters's Thesis of C. Reiling \citep{Reiling.2019}, which provides the groundwork for the simulation of the relevant physical processes inside the source.

\section{The experiment}
\label{ch:katrin}

At the KATRIN experiment, the neutrino mass is inferred from the measurement and subsequent spectral shape analysis of tritium $\upbeta$-decay electrons \citep{Aker.1132021}. The experiment has taken data since 2018. The data of the first two measurement campaigns combined yielded the world's best neutrino mass limit from $\upbeta$-spectroscopy with \SI{0.8}{\eV} (90\% CL) \citep{aker2021direct}. The target sensitivity will be reached after five years of operation \citep{KATRIN2005}, reducing the statistical uncertainty by a factor of approximately ten to the previous result by Kraus et al. \citep{kraus.2005}.

\subsection{Overview and principle of work}
\label{sec:kat:over}
\begin{figure}[tbp]
	\centering
	\includegraphics[width=1\linewidth]{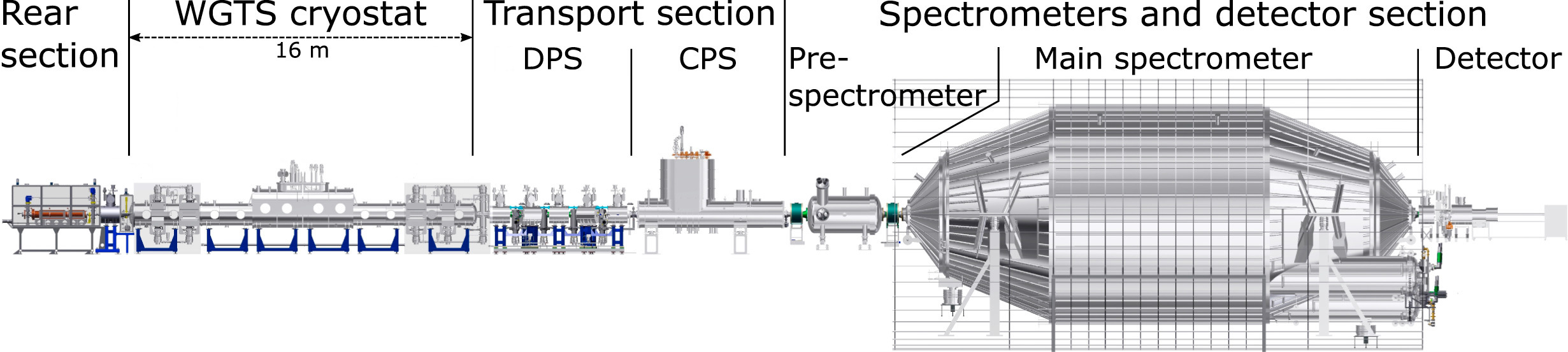}
	\caption{Overview of the beam line. Tritium gas is injected in the center of the windowless gaseous tritium source (WGTS) cryostat and pumped off at both sides. The neutral gas flow is further reduced in the differential pumping section (DPS) and the cryogenic pumping section (CPS). The simulations presented below will cover the \SI{16}{m} long WGTS cryostat, where the creation and interaction probability of electrons and ions is the highest. Figure adapted from \citep{PhDHeizmann2019}.}
	\label{fig:kat:overview}
\end{figure}

The KATRIN experiment uses single $\upbeta$-spectroscopy of molecular tritium \ch{T2} to determine the electron neutrino mass. The experimental setup is shown in figure~\ref{fig:kat:overview}. It consists of seven main components, which will be described in the following.

In the Windowless Gaseous Tritium Source (WGTS) $\upbeta$-electrons are created by $\upbeta$-decay of \ch{T2}, with an endpoint energy of $E_0 \approx \SI{18.6}{\kilo \eV}$ \citep{Otten2008}. Molecular tritium gas is continuously injected in the center of the source and differentially pumped-off at both ends using turbo molecular pumps (TMP) \citep{Kuckert.2018}. The gas decays with an activity of \SI{{}e11}{\becquerel} \citep{KATRIN2005}. Strong magnetic fields guide the electrons along the beam tube \citep{Arenz.2018b}. ''At the upstream end, the WGTS terminates in a gold-plated rear wall''\citep{Aker.1132021}. At the downstream end, the WGTS is connected to the differential pumping section \citep{Marsteller.2021} and the cryogenic pumping section \citep{Gil.2017}. These two sections further reduce the flux of remaining neutral tritium by a factor of at least \num{E15} \citep{Aker.352021}. This value exceeds the design requirement of \num{E14}, posted by consideration of residual tritium background in the spectrometers \citep{KATRIN2005}. The high energy electrons, which leave the pumping sections, are then filtered at the spectrometers. The spectrometers use the MAC-E filter principle \citep{Aker.352021}. The transversal kinetic energy of the electrons is converted to longitudinal kinetic energy. The movement is opposed by an electric potential of the spectrometers. Only the small fraction of electrons with an energy greater than the electric potential can pass and are measured at the detector. The neutrino mass is then inferred from the shape of the measured spectrum.

In total, most of the electrons leaving the WGTS in the direction of the DPS are either reflected by the large potential of the spectrometers or are reflected by the relatively small potential of the dipole electrodes of the DPS, which are installed in the DPS to filter out the ions through the use of the $E \times B$ drift \citep{Aker.352021}.

\begin{figure}[tb]
	\centering
	\includegraphics[width=0.7\linewidth]{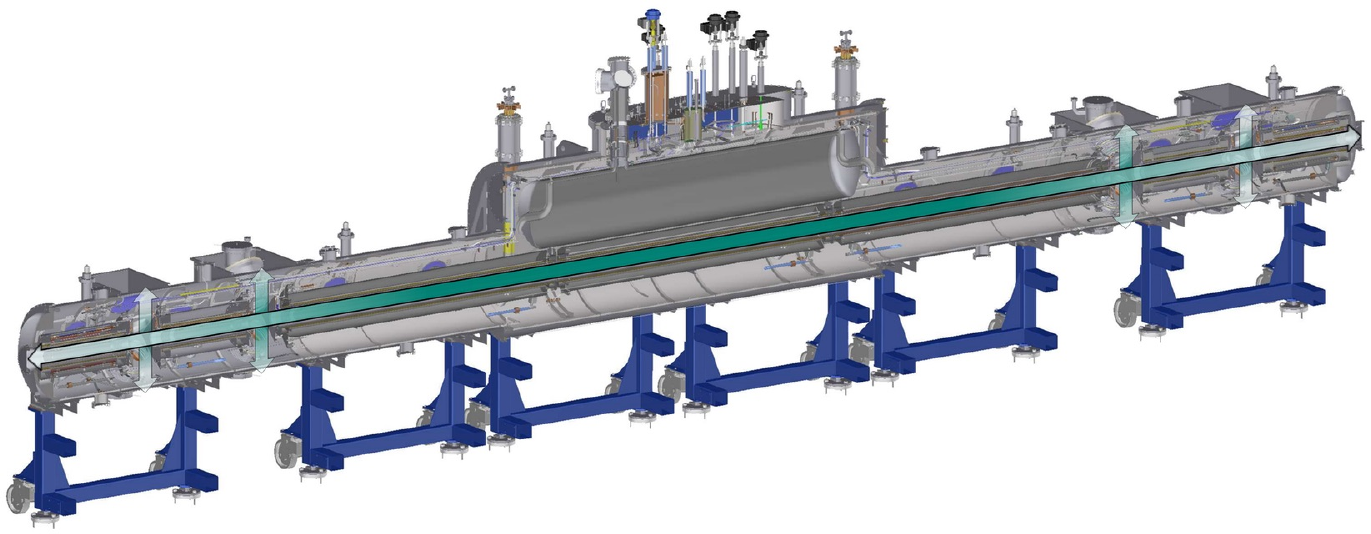}
	\caption{''Cross-sectional view of the WGTS. The gas density profile inside the beam tube is shown in green'' \citep{Heizmann:2017aa}, with flow direction indicated by the arrows. The neutral gas flow, which will be obtained through simulations by Kuckert et al. \citep{Kuckert.2018}, will be used as an input for our simulations presented below. Figure from \citep{Heizmann:2017aa}.}
	\label{fig:kat:wgts}
\end{figure}

The WGTS consists of a cylindrical source tube, with a length of \SI{16}{\meter} and a radius of \SI{4.5}{\centi\meter}, see figure~\ref{fig:kat:wgts}. Tritium is injected in the center, with a stable injection pressure of up to \SI{{}e-3}{\milli \bar} \citep{Aker.352021}. The gas then streams to both sides of the inlet and is mostly pumped out at the four attached pumping ports through turbomolecular pumps at approximately \SI{5}{\meter} and \SI{6.5}{\meter} distance from the inlet. In total, the tritium density drops from the center of the WGTS towards its sides by a factor of approximately \num{1000} \citep{Kuckert.2018}. The pumped gas is purified and ultimately fed back into the source. Overall, the WGTS can produce a column density of \SI{5e21}{\per \square \meter} for standard KATRIN operation \citep{Aker.352021} with a stability of \num{1} part in \num{1000} \citep{aker2021direct}.

The WGTS is cooled down to reduce thermal Doppler blurring of the $\upbeta$-spectrum ''and to ensure a high tritium density at low flow rate. On the other hand, a low tritium pressure and small clustering of tritium molecules is required'' \citep{PhDMachatschek2021}. The measurement campaigns KNM1 and KNM2 were therefore performed with a \SI{30}{\kelvin} setting \citep{aker2021direct, Aker:2019aa}. All following campaigns are performed at \SI{80}{\kelvin} ''to allow for equal source conditions in tritium and krypton-83m commissioning measurements''~\citep{PhDMachatschek2021}. Thus, the results presented here were calculated for the \SI{80}{\kelvin} setting.

The WGTS cryostat also houses seven superconducting magnets, which provide the constant guiding magnetic field of \SI{2.5}{\tesla} \citep{Heizmann:2017aa} for the electrons.

\subsection{Plasma effects in the WGTS}
\label{sec:kat:plasma}
The final neutrino mass sensitivity of \SI{0.2}{\eV} can only be reached through a precise examination of all systematic effects \citep{Aker.1132021}. One of these effects is the plasma inside the source. It is generated through $\upbeta$-decay and subsequent impact ionization of neutral tritium gas (all relevant processes are listed in section~\ref{sec:physprocesses}). The resulting charged particle density is sufficient to assume a plasma inside the WGTS \citep{Kuckert.2016}. The effects of this plasma can influence the motion of the electrons.

The plasma, in conjunction with boundaries, exerts a potential throughout the source, also called starting potential, which can show a non-homogeneous structure. The final energy of the $\upbeta$-electrons selected by the spectrometers can therefore be dependent on the original position of $\upbeta$-decay inside the source. Hence, the plasma potential can distort the integral spectrum measured at the detector, and has to be accounted for in the analysis by a quantitative understanding of the plasma. Only the inhomogeneous part of the potential is of special interest, as the homogeneous part shifts the entire $\upbeta$-spectrum and can be easily taken into account.

In the past, the diffusion ansatz was used by L. Kuckert \citep{Kuckert.2016} to provide a quantitative model of the plasma. Recent studies indicate, that this ansatz may not include all significant plasma effects, such as the high energy tail of the spectrum, the transition between a collisional and non-collisional plasma, as well as all time dependent effects. Therefore, a new kinetic simulation is being developed within the KATRIN collaboration. A precondition for any theoretical plasma study is the knowledge of electron and ion distributions (and velocities) at any point within the tritium source. In the following sections, we will describe how these density and velocity distributions can be calculated.

\section{Simulation approach}

Electrons within the WGTS are generated through $\upbeta$-decay and through subsequent secondary processes. For each $\upbeta$-decay, this forms a Markov chain of events, with further branching event chains for secondary particles. This scenario is best simulated using the Monte Carlo technique.

The Monte Carlo algorithm was implemented in a newly developed code, named KARL. The code base for KARL was initiated by C. Reiling \citep{Reiling.2019}, but significant changes were made since then and erroneous calculations corrected. The algorithm of the new code version is described in section~\ref{sec:montec}. The physics is outlined in section~\ref{sec:physprocesses} and the first results presented in section~\ref{sec:results}.

\subsection{Monte Carlo scheme}
\label{sec:montec}

\begin{figure}[tb]
	\centering
	\pgfimage{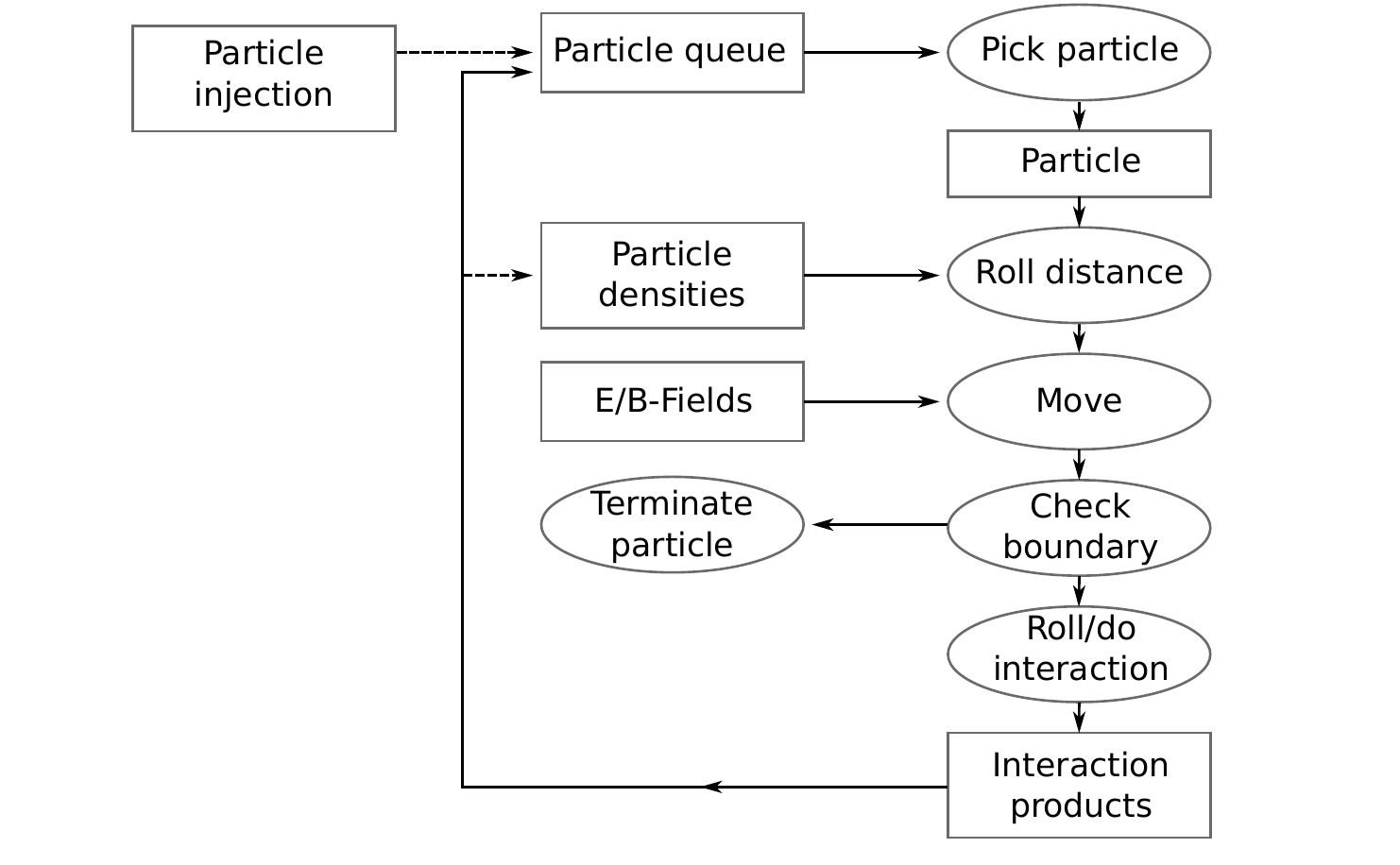}
	\caption{Monte Carlo scheme of the KARL code. $\upbeta$-electrons and ions are added to the particle queue. A particle is picked from the queue and processed. The iteration step either ends in a particle interaction or through termination at the simulation boundaries. New $\upbeta$-decay particles are added when the particle queue is empty. The tritium density is specified before the simulation. The particle densities of electrons and ions are determined during the simulation.}
	\label{fig:mon:flow_diagram}
\end{figure}

Each particle chain starts at the $\upbeta$-decay, where the position of the new electron and corresponding ion is sampled from the known neutral particle density distribution \citep{Kuckert.2018}. The energy is sampled from the Fermi spectrum of the $\upbeta$-decay \citep{Fermi.1934}. The $\upbeta$-decay particles are then added to the particle queue and the simulation starts its iterative cycle, see figure~\ref{fig:mon:flow_diagram}. In each iteration step, a particle is selected at random from the particle queue. Then the mean free path, see equation~\eqref{eq:mfp}, is determined from all possible interacting partner densities at the position of the particle and the energy of the particle. The particle is then moved according to the mean free path and the electromagnetic fields to a new position, see section~\ref{ssec:charged_particle_motion}. If the particle hits the simulation boundaries, it will be deleted. Otherwise, an interaction is selected randomly from the available interactions, see section~\ref{sec:mon:interactions}. The interaction products are then added to the particle queue and the iteration starts from the beginning. The particle loop is repeated until there are no more particles in the particle queue. Then, new $\upbeta$-decay particles are created and processed until the specified target decay number is reached. The particles of the simulation can be used to determine particle densities during the simulation, see section~\ref{sec:mon:density_fields}. These densities can then be used to simulate electron-ion interactions.

\subsection{Particle interactions}
\label{sec:mon:interactions}

\subsubsection{Two particle scattering}
\label{ssec:mon:interactions:two_particl_scattering}

In the simulation, particles interact with other particles. The probability of this interaction is given by the total cross section (TCS). In all cases presented here, it is dependent on the energy $E$ of the interacting partners. In the following, the TCS is always given in the inertial frame of reference, where the target particle is at rest. The probability for an interaction $w$ of a particle after passing a distance $x$ assuming a constant density of target particles $n_\mathrm{T}$ is given by
\begin{equation}
	\label{eq:mon:prob}
	w = \sigma(E) \cdot n_\mathrm{T} \cdot x\, .
\end{equation}
Eq.~\eqref{eq:mon:prob} can be used to determine the average distance between two interactions, called mean free path. It calculates as
\begin{equation}
	\lambda(E) = \frac{1}{\sigma(E) \cdot n_\mathrm{T}} \, . \label{eq:mfp}
\end{equation}
This equation needs to be adapted if more than one interaction channel exists. For this reason, a new quantity, named the total macroscopic cross section (TMCS), is defined through
\begin{equation}
	\label{eq:mon:tmcs}
	\Sigma_\mathrm{tot}(E) = \sum_i \Sigma_i(E) \, ,
\end{equation}
where $\Sigma_i(E) = \sigma_\mathrm{i}(E) \cdot n_\mathrm{T,i}$ is named the macroscopic cross section (MCS) of the interaction channel $i$. $n_\mathrm{T,i}$ is the corresponding target particle density. The mean free path then calculates to
\begin{equation}
	\label{eq:mon:lambda}
	\lambda_\mathrm{tot}(E) = \dfrac{1}{\Sigma_\mathrm{tot}(E)} \, ,
\end{equation}
rendering the TMCS a direct indicator for the mean free path.

After interaction, the final state of a specific property of the particles can be important for the simulation. The statistical distribution of the final state $\zeta$ is given by the differential cross section (DCS)~$\frac{\upd \sigma}{\upd \zeta}$. The probability for observing a particle within the parameter space between $[a,b]$ is given by
\begin{equation}
	w = \frac{1}{\sigma} \cdot \int_a^b \frac{\upd \sigma}{\upd \zeta} \upd \zeta \, .
\end{equation}
In the context of this work, the relevant properties that define the final state are the energy of the particles after collision and their scattering angle.

\subsubsection{Reaction rates}
The cross section of some reactions is not available from the literature. This occurs mostly for rare and low energetic interactions. In these cases, reaction rate coefficients $k$ can be used to provide a rough estimate of the cross section. For an unimolecular elementary step, the reaction rate $r$ is calculated by
\begin{equation}
	r = k \cdot n \, ,
\end{equation}
where $n$ is the initial density of particles. From a microscopic point of view, this rate can be identified with the collision rate $\nu$
\begin{equation}
	r = \nu = \frac{\bar{\mathrm{v}}}{\lambda} \, ,
\end{equation}
where $\bar{\mathrm{v}}$ is the mean velocity of the interacting particle and $\lambda$ the mean free path. The velocity can be calculated from the kinetic energy $E$ of the interacting particle. The mean free path is calculated using equation \eqref{eq:mfp}. In total, the cross section can be written as
\begin{equation}
	\sigma = \frac{k}{\sqrt{\frac{2 E}{m}}} \, ,
	\label{eq:rate_to_cross_section}
\end{equation}
where $m$ is the mass of the interacting particle. Thus, the reaction rates coefficients can be used as an approximate estimate of the energy dependency of the cross section.

\subsection{Charged particle motion}
\label{ssec:charged_particle_motion}
The particle motion is bound by the mean free path and the electromagnetic background fields in the source. The actual traveled distance $L$ is determined in the simulation through a pseudo random number $\eta \in (0,1)$ and the mean free path $\lambda$
\begin{equation}
	L = - \lambda \cdot \ln \eta \, ,
\end{equation}
taking into account the randomness of the real motion. The magnetic field is assumed to be temporally and spatially homogeneous. Therefore, the movement can be performed by the drift approximation \citep{Chen.1984}. In the drift approximation algorithm, the movement of the particle is described by the superposition of the movement of the guiding center and the gyro motion of the particle around the guiding center.

The movement of the particle is subdivided into small spatial steps $\Delta x$ to account for changes in the electromagnetic field during movement. The step size is chosen by the step size of the electromagnetic field data, to avoid movement across multiple electric field cells. The discretization of the spatial dimension can be translated into a discretization of the time domain to
\begin{equation}
	\Delta t = \frac{\Delta x}{|\vec{v}^{i}|} \, ,
\end{equation}
where $|\vec{v}^i|$ is the absolute velocity of the particle before movement. The movement algorithm is performed in three steps. First, the position $z$ in longitudinal direction is calculated by
\begin{equation}
	z^{i+1} = z^{i} + v_\parallel^i \Delta t \, ,
\end{equation}
where the index $i$ denotes the timestep at which the quantity is given. $v_\parallel^i$ is the velocity of the particle parallel to the magnetic field. Second, the electric field during the motion is approximated by the mean of the field value at the original position and at the new position. This field is then used to calculate the movement perpendicular to the magnetic field and used to determine the new velocity $v_\parallel^{i+1}$. The perpendicular particle position changes to
\begin{eqnarray}
	x^{i+1} =& x^{i} + v_{\perp,x} \Delta t \\
	y^{i+1} =& y^{i} + v_{\perp,y} \Delta t \, ,
\end{eqnarray}
where $v_\perp$ is given by the so called E-cross-B drift \citep{Chen.1984}
\begin{equation}
	\vec{v}_\perp = \frac{\vec{E} \times \vec{B}}{B^2} \, .
\end{equation}
Third, the parallel velocity changes corresponding to the Lorentz force to
\begin{equation}
	v_\parallel^{i+1} = v_\parallel^i + \frac{q E_\parallel}{m} \Delta t \, ,
\end{equation}
where $E_\parallel$ is the electric field strength parallel to the magnetic field, $m$ is the mass of the particle and $q$ its charge.

The movement of a particle is stopped, either when the particle reaches an absorbing surface or if it crosses a density bin border, see section~\ref{sec:mon:density_fields}. If the particle movement is stopped at a density bin border, the movement step is performed again from the beginning. This method ensures a more precise description of particle movement through a cloud of particles with varying density. The absorbing surfaces for particle deletion depend on the particle type, see section~\ref{sec:kat:over}. Ions, on the one hand, are deleted each time they hit the beam tube walls, the rear wall, or leave the source at the DPS side. The latter represents the filtering of the ions by the dipole electrodes in the DPS. Electrons, on the other hand, are deleted, when they hit the beam tube walls or the rear wall. They are reflected back to the simulation domain, when they cross the boundary towards the DPS. This procedure represents the reflection of the electrons both at the DPS dipole electrodes or the spectrometer potential. Backscattering of electrons and ions at the metal surfaces is not considered in the current version of the code.

\subsection{Secondary particles as target field}
\label{sec:mon:density_fields}
Electrons and ions mostly interact with the neutral gas in the source due to the large neutral particle density. These interactions depend solely on the primary particle and the neutral gas properties (density, velocity distribution). The gas density is high enough that it can be assumed that the neutral gas flow is not influenced by charged particle interactions \citep{Aker.352021}. Thus, each Monte Carlo event would be independent of the other, which is common for most Monte Carlo codes. Nevertheless, electrons and ions can interact with each other in the source, see section~\ref{sec:physprocesses}. Therefore, each Monte Carlo event must be dependent on the other events. This dependence is mimicked in the simulation by so-called density fields. These density fields determine the number density of each particle specie during the simulation. These fields are then used to determine the TMCS, see equation~\eqref{eq:mon:tmcs}.

\begin{figure}[tb]
	\centering
	\pgfimage{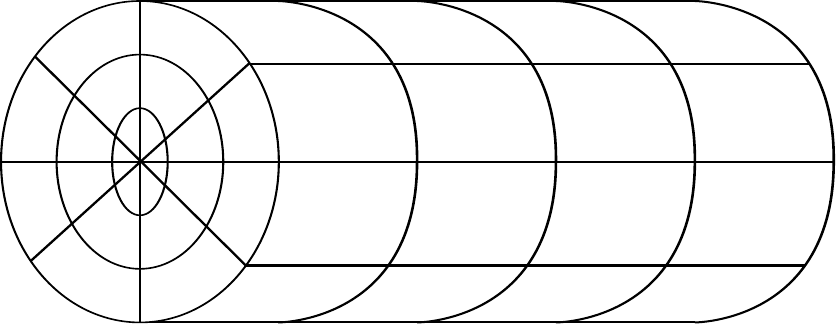}
	\caption{Schematic representation of the source segmentation for density fields and particle current calculation. The source is segmented into evenly spaced segments in longitudinal, radial and azimuthal direction predefined by the user. The segment boundaries are used as predefined planes for the determination of particle currents, see section~\ref{ssec:particle_tracking_and_diagnostics}.}
	\label{fig:source_segmentation}
\end{figure}

In the simulation, the source is segmented into evenly spaced segments in longitudinal, radial and azimuthal direction, predefined by the user, see figure~\ref{fig:source_segmentation} for a schematic representation. In each iteration step of the particle queue, it is recorded how long a particle with index $i$ stays in one of the predefined segments, labeled by the time $t_i$. The number density $n_\alpha$ of a species $\alpha$ is then proportional to the total time of all particles of the corresponding species spent in the segment $\sum_i t_i$ and the volume $V$ of the segment. In case of recombination, an additional term has to be added corresponding to the number of recombination $N_\gamma$ of the recombination partner $\gamma$ in the segment, resulting in
\begin{equation}
	n_\alpha = \frac{\sum_i t_i}{t_\text{sim} V} - \frac{N_\gamma}{V}\, ,
	\label{eq:target_field_density}
\end{equation}
where $t_\text{sim}$ is called the simulated time (formula adapted from \citep{Reiling.2019}). It describes the simulated physical time that is past after simulating a fixed number of beta decays $N_\beta$. It can therefore be directly determined by the activity $a$ of the source through
\begin{equation}
	t_\text{sim} = \frac{N_\beta}{a}\, . \label{eq:simulated_time}
\end{equation}
The activity of the source is not dependent on the interaction of the charged particles in the source, and can therefore be evaluated from the predefined amount of neutral gas in the source.

The number density $n_\alpha$ can also be determined depending on the energy of the particle. These number densities are then binned to a predefined energy scale, resulting in the spectrum of the particles. Thus, the density fields can be used in the calculation of the TMCSs, see equation~\eqref{eq:mon:tmcs}, and also used for sampling a new particle with the simulated spectrum. The sampled particles can then be used in the interaction step, see section~\ref{sec:physprocesses}, to evaluate the outcome of a particle interaction of two different types of particles. Therefore, these new sampled particles are called interaction partners.

The sampling of an interaction partner is performed in three different steps: sample particle energy, roll velocity direction, copy position of primary particle. First, the energy of the interaction partner is sampled from the energy distribution of the corresponding density field at the position of the primary particle. Secondly, a velocity direction is picked at random. The velocity of the new particle is then scaled according to the energy and mass of the particle. Finally, the new particle is generated with this velocity and the position of the primary particle.

\subsection{Particle tracking and diagnostics}
\label{ssec:particle_tracking_and_diagnostics}

The Monte Carlo approach of the simulation needs high enough particle statistics. The statistic is determined by the number of generated $\upbeta$-decay events $N_\beta$, specified by the user. Each of these events produces multiple secondary particles. Recording of each position and particle would result in too much data. Thus, it was decided on three different output methods: density fields, see previous section~\ref{sec:mon:density_fields}, absolute particle numbers of conversion and termination, and particle currents through virtual barriers.

The density fields with number densities $n_\alpha$ of the previous section are recorded each time after a specified number of generated $\upbeta$-decay events. The last output is performed after $N_\beta$ decays. This output is used for the final determination of the particle number densities of the simulation, which is one of the main objectives of the code. The other outputs are only used for diagnostics of the code.

Special counters are implemented to track each time a particle is terminated at the simulation boundary, converted to another particle species or recombines. The termination counter is selective on the boundary type: rear wall, tube wall or DPS. These counters can be used for diagnostics of the code or for tests of the code.

The second main objective of the code is to determine the particle currents in the source. This task is performed through so-called virtual barriers: predefined planes within the source, where each particle is registered, when it crosses from one side to another. The total current through the segment faces is then given by
\begin{equation}
	j = \frac{N^+ - N^-}{A \, t_\text{sim}} \, ,
\end{equation}
(formula from \citep{Reiling.2019}) where $N^+$ and $N^-$ is the number of particles crossing the faces in positive and negative direction, $A$ is the area of the face and $t_\text{sim}$ the simulated time, see~Eq.~\ref{eq:simulated_time}. For convenience, the location of the virtual barriers are chosen to coincide with the segment boundaries, see figure~\ref{fig:source_segmentation}, of the density field calculation, see section~\ref{sec:mon:density_fields}.

The current through the faces can be determined depending on the energy and type of species crossing the faces. This is performed through a selective binning to a predefined energy scale during the simulation. The absolute current is then given by the sum over all energy bins.

The accuracy of the simulation depends on the number of decays $N_\beta$. This number is inserted directly in the calculation of the current and the density, but it also contributes indirectly through the production of secondary particles. The specific contributions are difficult to calculate directly because of the multitude of different interactions. Hence, the accuracy of the simulation was determined through multiple simulations with the same number of decays but with different random number generator seeds. The mean of the normalized standard deviation of each energy bin was calculated. For a counting experiment, it is expected that the standard deviation scales with $1/\sqrt{N_\beta}$, which was found to be in good agreement with our test simulation results. To reduce the statistical uncertainty to \SI{0.1}{\percent}, it was decided to use \num{1E6} $\upbeta$-decays in the following simulations.

\section{Physical processes}
\label{sec:physprocesses}
In this section, the most relevant physical processes for charged particles in the source are described. The focus lies here on the implementation in the code. For the interactions, the cross sections were calculated by analytical formulas or through numerical tables. The corresponding parameters and values were obtained from the literature. Initially, this collection was compiled by C. Reiling \citep{Reiling.2019}.

Due to the nature of the Monte Carlo approach, uncertainties of the cross sections can only be considered in the code through an evaluation of multiple simulations with varied cross sections, which will be done in the future. The interested reader is referred to the original publications for the evaluation of the uncertainties for the parameters.

\subsection{Particle injection}

The injection of particles mimics the $\upbeta$-decay of the tritium molecules in the source into an electron \ch{e-} and an ion molecule \ch{T-He+}. In the code, the ion is described as a \ch{T2+} ion. Isotopic differences are assumed to be small \citep{schimpf.2021}. Additionally, due to charge transfer processes from \ch{T-He+} to \ch{T-T+} and dissociation, the lifetime of \ch{T-He+} is negligible in this context \citep{Bodine.2015}. The position of the new $\upbeta$-electron and ion is sampled from the known density distribution of the neutral gas through the acceptance-rejection method \citep{Casella.2004}. The energy of the $\upbeta$-electron is sampled through the Fermi distribution of $\upbeta$-decay \citep{Fermi.1934} in conjunction with the acceptance-rejection method. The velocity direction is assumed to be isotropically distributed. The energy of the ion is sampled from a Maxwell-Boltzmann distribution with an additive drift velocity of the neutral gas. The recoil of the ion is neglected due to the large mass difference.

\subsection{Interactions of electrons with tritium molecules} \label{sec:mod:eT2}

$\upbeta$-electrons are created in the source with energies of up to \SI{18.6}{\kilo \eV}. In contrast, the molecular tritium gas has a temperature of \SI{80}{\kelvin}, which corresponds to energies of \SI{6.9}{\milli \eV}. Thus, \ch{e-}-\ch{T2} interactions need to be described on a large energy range. Density-wise, the WGTS is dominated by molecular tritium. Hence, interactions with molecular tritium are very probable, and the possible interactions need to be modeled precisely.

\begin{figure}[tb]
	\centering
	\pgfimage{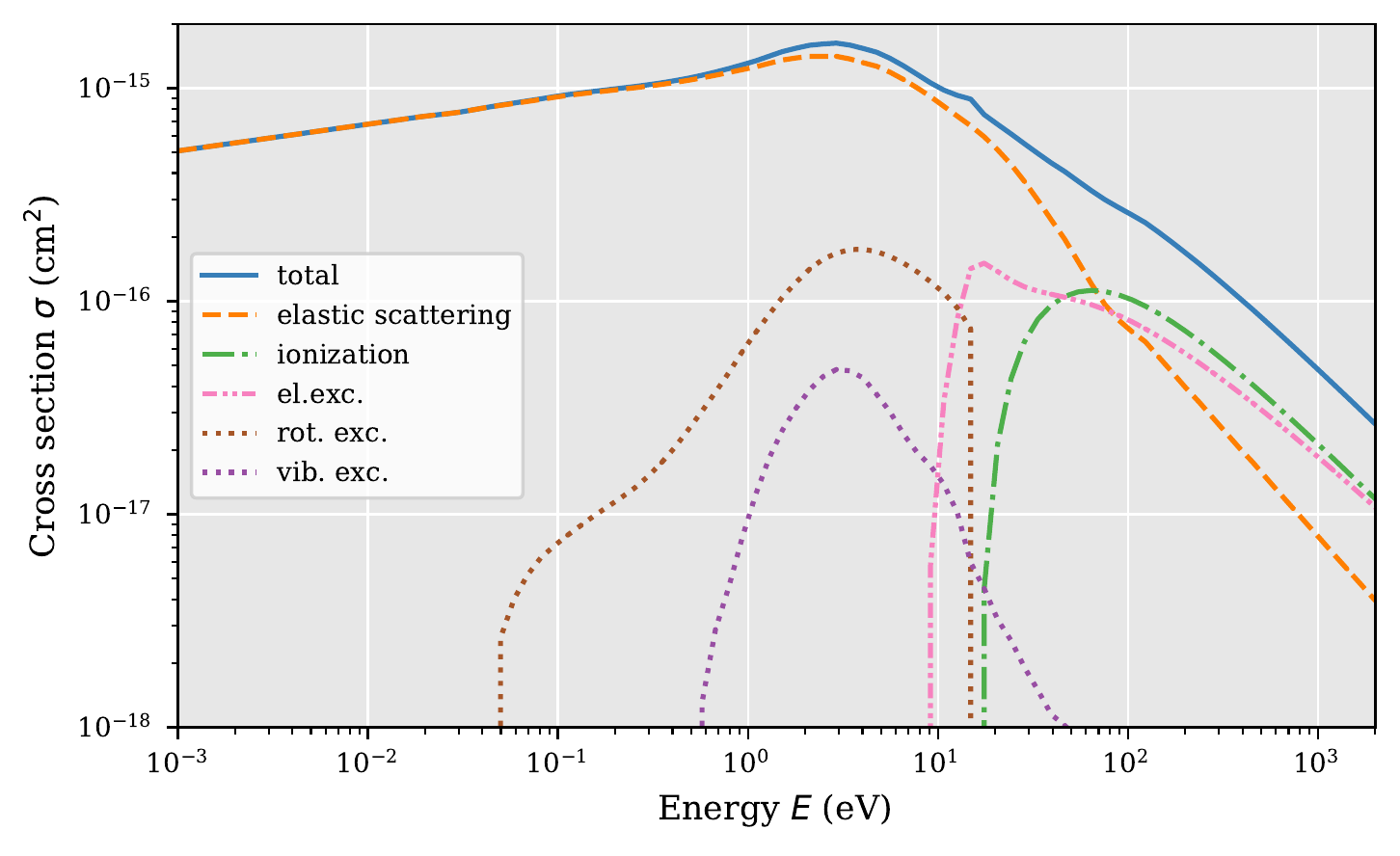}
	\caption{Overview of implemented \ch{e-}-\ch{T2} interactions. Displayed ionization and excitation cross sections are summed over all sub-channels. Energies are defined in the laboratory frame, in which kinetic electrons interact with fixed tritium molecules. Data for rotational excitation, vibrational excitation taken from \citep{Yoon:2008aa}. Data for ionization and electronic excitation taken from \citep{Janev:2003aa}. Data of elastic scattering from \citep{Yoon:2008aa}, simulation using ELSEPA \citep{Salvat:2005aa} and extrapolation, see below. Figure based on \citep{Reiling.2019}.}
	\label{fig:mod:eT2_overview}
\end{figure}

A collection of the implemented models for \ch{e-}-\ch{T2} interactions is shown in the following sections, see also figure ~\ref{fig:mod:eT2_overview}. There is almost no data available for interactions with tritium. Isotopic differences of the interaction probability are assumed to be small \citep{schimpf.2021}. Therefore, models of \ch{e-}-\ch{H2} interactions are adopted in the code.

\subsubsection*{\ch{e-} - \ch{T2} elastic scattering}
\label{sssec:mod:eT2:elastic}

Elastic scattering with molecular tritium is the dominant interaction channel for low energy electrons ($E < \SI{60}{eV}$), see also figure~\ref{fig:mod:eT2_overview}. For energies above $\SI{60}{eV}$, other interactions, such as ionization and electronic excitation, become dominant. Nevertheless, elastic scattering has still an effect on the particle motion.

\begin{figure}[tb]
	\centering
	\pgfimage{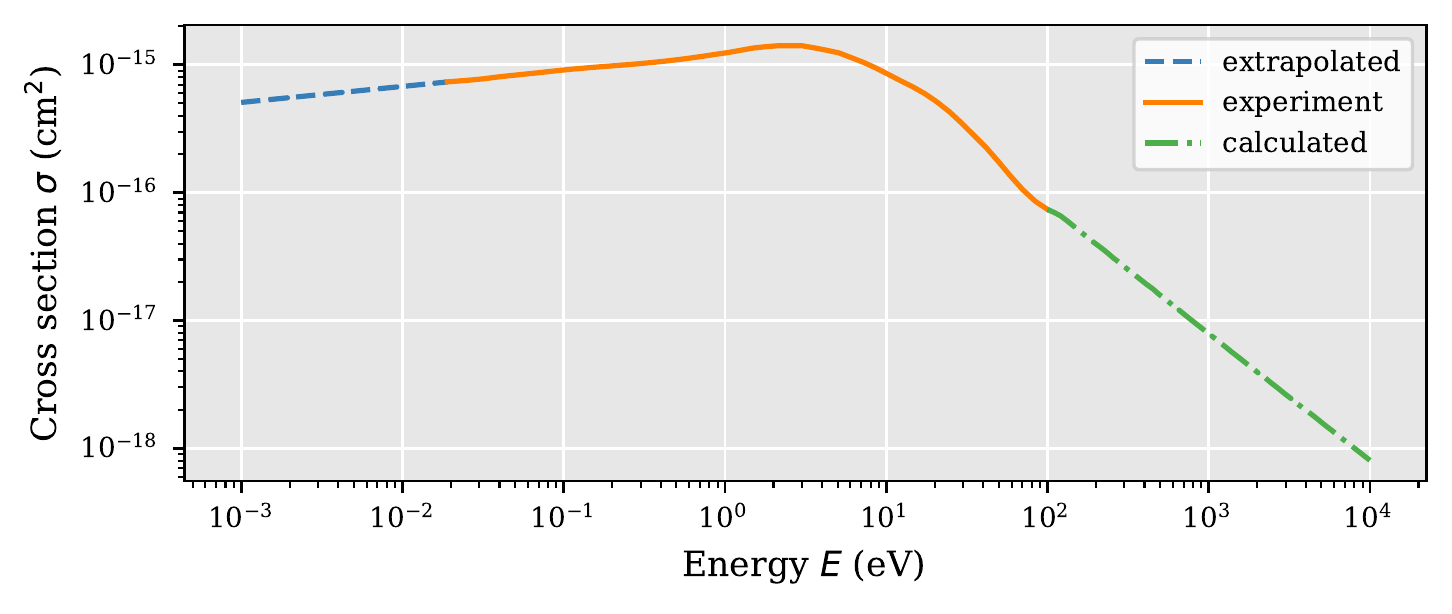}
	\caption{Visualization of the composition of the cross section of \ch{e-}-\ch{T2} elastic scattering. The cross section is derived from three models. Cross sections for energies $E > \SI{100}{ \eV}$ were calculated using ELSEPA \citep{Salvat:2005aa}. Cross sections for energies in the range $\SI{0.02}{eV} < E < \SI{100}{eV}$ were adapted from experimental data from \citep{Yoon:2008aa}.
		Cross sections for energies $E < \SI{0.02}{eV}$ are extrapolated values from experimental data from \citep{Yoon:2008aa}. Figure based on \citep{Reiling.2019}.}
	\label{fig:mod:eT2_elastic}
\end{figure}

\begin{figure}[tb]
	\centering
	\pgfimage{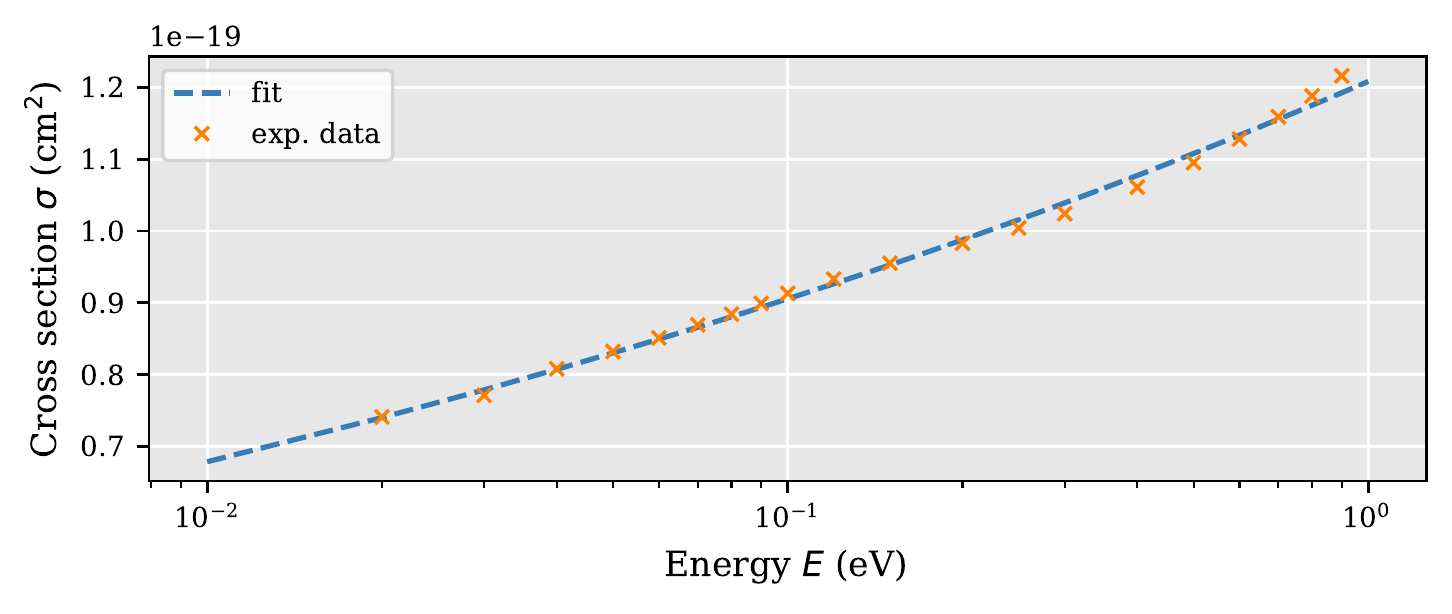}
	\caption{Exponential fit of \ch{e-}-\ch{T2} elastic scattering cross section data below \SI{1}{\eV} from \citep{Yoon:2008aa}. The resulting fit function is used for cross sections below \SI{0.02}{\eV}. Figure based on \citep{Reiling.2019}.}
	\label{fig:app:eT2_ela_lr}
\end{figure}

There was no model available for the cross section of elastic scattering on the large energy range. Therefore, the TCS was derived from three different sources, see also figure~\ref{fig:mod:eT2_elastic}. The majority of the energy range ($\SI{0.02}{eV} < E < \SI{100}{eV}$) was covered by the experimental studies from \citep{Yoon:2008aa} and implemented as an analytical function. The cross section for higher energies $E > \SI{100}{eV}$ were derived from the ELSEPA simulation code \citep{Salvat:2005aa} and implemented as numerical tables. No experimental or theoretical cross sections were found for energies below $E < \SI{0.02}{eV}$. The shape of the cross section of \citep{Yoon:2008aa} indicates a continuous trend towards lower energies. Thus, the cross sections of Yoon was extrapolated towards lower energies as a first approximation, see also figure~\ref{fig:app:eT2_ela_lr}. The analytical expression of this extrapolation (from \citep{Reiling.2019}) reads as
\begin{equation}
	\sigma (E) = \SI{1.21e-19}{\square \meter} \cdot \Big ( \frac{E}{\si{\eV}} \Big )^{0.125} \, .
\end{equation}

The final state of the elastically scattered electrons depends on the two scattering angles $\varphi$ and $\theta$. The orientation of the tritium molecules is assumed to be isotropically distributed. Thus, the polar angle is sampled from the interval $[0, 2 \pi)$. The distribution of the azimuthal angle is determined through the use of the ELSEPA code \citep{Salvat:2005aa}. The simulations showed, that the azimuthal angle can be assumed to be isotropically distributed for energies below $E < \SI{10}{eV}$. For higher energies, the azimuthal angle is sampled from the tabulated data of the ELSEPA calculations.

The position of the electron does not change in the scattering process. Nevertheless, the guiding center of the gyro motion changes its positions. Thus, this change of the guiding center is reflected in the simulation. The new position of the guiding center is shifted by the length $\Delta x$ with
\begin{equation}
	\label{eq:position_change_el_scattering}
	\Delta x = 2 r_L \sin \left(\frac{\theta}{2}\right) \, ,
\end{equation}
where $\theta$ is the projected scattering angle to the magnetic field direction. The electron performs many gyrations between interactions. Therefore, the polar angle of the gyration can be considered as randomly distributed. Thus, the shift is applied randomly in the x and y directions.

\subsubsection*{Electron impact ionization of \ch{T2}}

\begin{figure}[tb]
	\centering
	\pgfimage{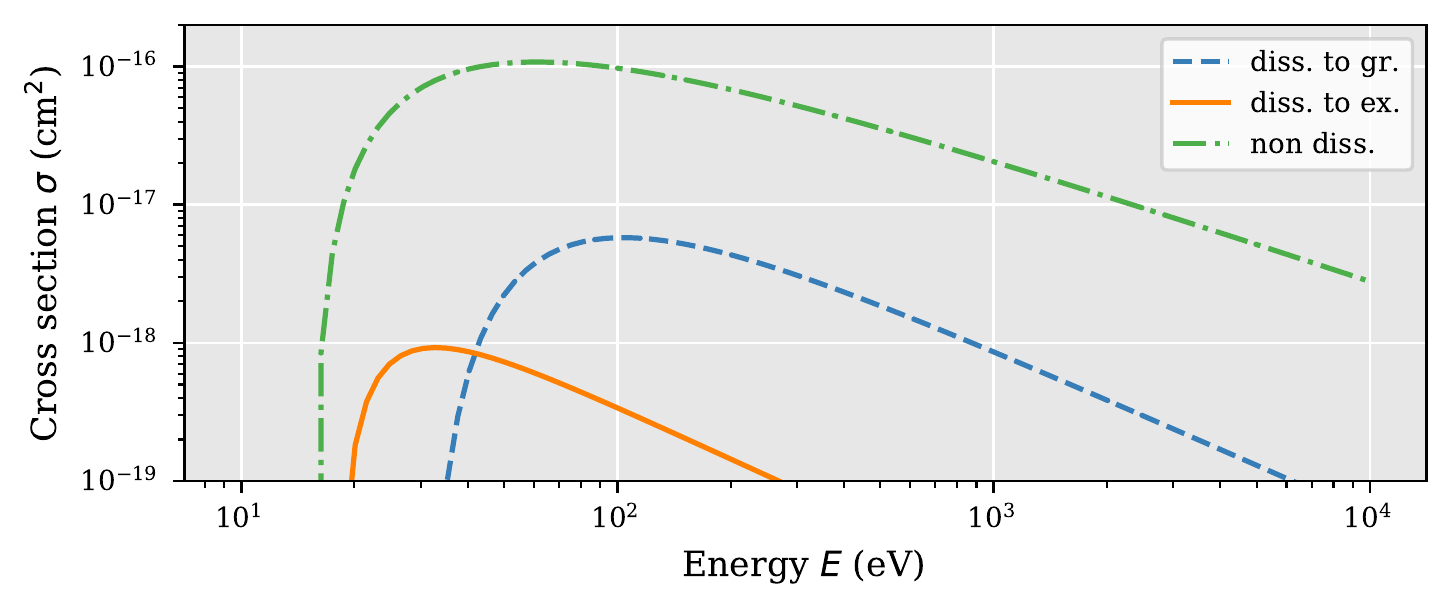}
	\caption{Cross section for electron impact ionization of \ch{T2} for the three different ionization channels: non-dissociative ionization (non diss.), dissociative ionization to the ground state (diss. to gr.) and dissociative ionization to the first excited state (diss. to ex.). Data from \citep{Janev:2003aa}. Figure based on \citep{Reiling.2019}.}
	\label{fig:app:eT2_ion}
\end{figure}

Most of the electrons and ions in the source are not created through $\upbeta$-decay, but through electron impact ionization. There exist multiple different ionization channels, which mostly depend on the state of the tritium molecule before, during and after ionization \citep{Janev:2003aa}. Because the tritium gas is cooled down to \SI{80}{K}, it is assumed here that the molecule is in the ground state, prior to the interaction. Similarly, it is assumed that the maximum reachable state of the molecule is the first excited state. In this case, the study of Janev et al. \citep{Janev:2003aa} provide analytical formula for three different ionization channels: non-dissociative ionization, and dissociative ionization over the ground and first excited state of the ion, see also figure~\ref{fig:app:eT2_ion}.

The most probable channel is non-dissociative ionization
\begin{equation}
	\ch{e- + T2 -> e- + T2+ + e-} \, ,
\end{equation}
with a threshold energy of \SI{15.42}{\eV}. Each electron creates an additional electron and a tritium ion molecule. In dissociative ionization, first an electron is removed from the molecule, which then resides in an excited state. This state then decays into a tritium ion and a tritium atom
\begin{equation}
	\ch{e- + T2 -> e- + T2^+^* + e- -> T+ + T + 2 e-} \, ,
\end{equation}
with threshold energies of \SI{18.15}{\eV} and \SI{30.6}{\eV} for the ground and the first excited state respectively.

In the simulation, the energy of the ions is assumed to follow the distribution of the gas flow. Therefore, the energy of the ion is sampled from the Maxwell-Boltzmann distribution of the neutral gas. The energy of the two final electrons is correlated. In the code, the energy of the first electron is sampled from the analytical formula of the DCS from \citep{Kim:1994aa}. The energy of the second electron is then calculated correspondingly as the difference between the initial energy, the ionization energy and the final energy of the first electron. The distribution of the azimuthal and polar angle of both electrons is adopted from \citep{Grosswendt:1978aa}.

\subsubsection*{Electronic excitation of \ch{T2} by \ch{e-}}
\label{sec:mod:electronic_excitation}

Tritium can be electrically excited by collision with electrons. The energy difference of the levels is on the order of $\mathcal{O}(\SI{10}{\eV})$. Thus, high energy electrons $E > \SI{100}{eV}$ can lose a significant amount of their energy due to excitation. Similar to electron impact ionization, the cross section of the interaction depends on the initial and final state of the molecule. Again, it is assumed, that the molecule is in the ground state prior to the interaction and that the maximal state of the molecule is the first excited state. In this case, the study of Janev et al. \citep{Janev:2003aa} provides three main categories of electronic excitation:
\begin{itemize}
	\item Excitation to dipole allowed singlet state
	\item Excitation to dipole forbidden singlet state
	\item Excitation to triplet states
\end{itemize}
Each of the categories can be subdivided by additional quantum numbers of the molecule. In the code, the cross section of the most dominant interactions from \citep{Janev:2003aa} were implemented as an analytical formula. This way, 15 different excitation channels were resolved.

In the simulation, it is assumed that the tritium molecule does not gain any additional kinetic energy due to the large mass difference. Thus, the electron only loses the energy of the excitation process. The direction of motion is therefore also left unchanged.

\subsection*{Rotational and vibrational excitation of \ch{T2} by \ch{e-}}
\label{sec:mod:rotation}

The energies of the rotational and vibrational excitation levels of \ch{T2} and \ch{H2} are assumed to be different, due to the significant mass difference of the isotopes. Studies of Yoon et al.\citep{Yoon:2008aa} and Janev et al. \citep{Janev:2003aa} show, that rotational excitation and vibrational excitation have a non-negligible effect on the total cross section. Therefore, in the code, the cross sections of \ch{H2} are used as a first approximation for rotational and vibrational excitation of \ch{T2}. The cross sections are implemented in the code as analytical formula, where the data of rotational excitation is taken from \citep{Yoon:2008aa} and the data of vibrational excitation is taken from \citep{Janev:2003aa}.

Similar to the electronic excitation, it is assumed that the tritium molecule does not gain any additional kinetic energy due to the large mass difference. Thus, the electron only looses the energy of the excitation process and the direction of motion is left unchanged.

\subsubsection*{Recombination of electrons with tritium ions} \label{sec:mod:eTI}

The direct interaction between ions and electrons is improbable, due to the expected low densities of electrons and ions ($n_e \approx n_i \approx \SI{E6}{cm^{-3}}$) \citep{Kuckert.2016}. Therefore, only recombination is considered in the code, because this reaction changes the total number of electrons and ions.

In the code, three different ion species are distinguished, see section~\ref{sec:mod:TIT2}. Therefore, only recombination processes with these species are discussed here. There was no data found on the recombination cross sections of tritium. Therefore, the cross section of recombination of tritium is approximated by the cross section of hydrogen.

\begin{figure}[tb]
	\centering
	\pgfimage{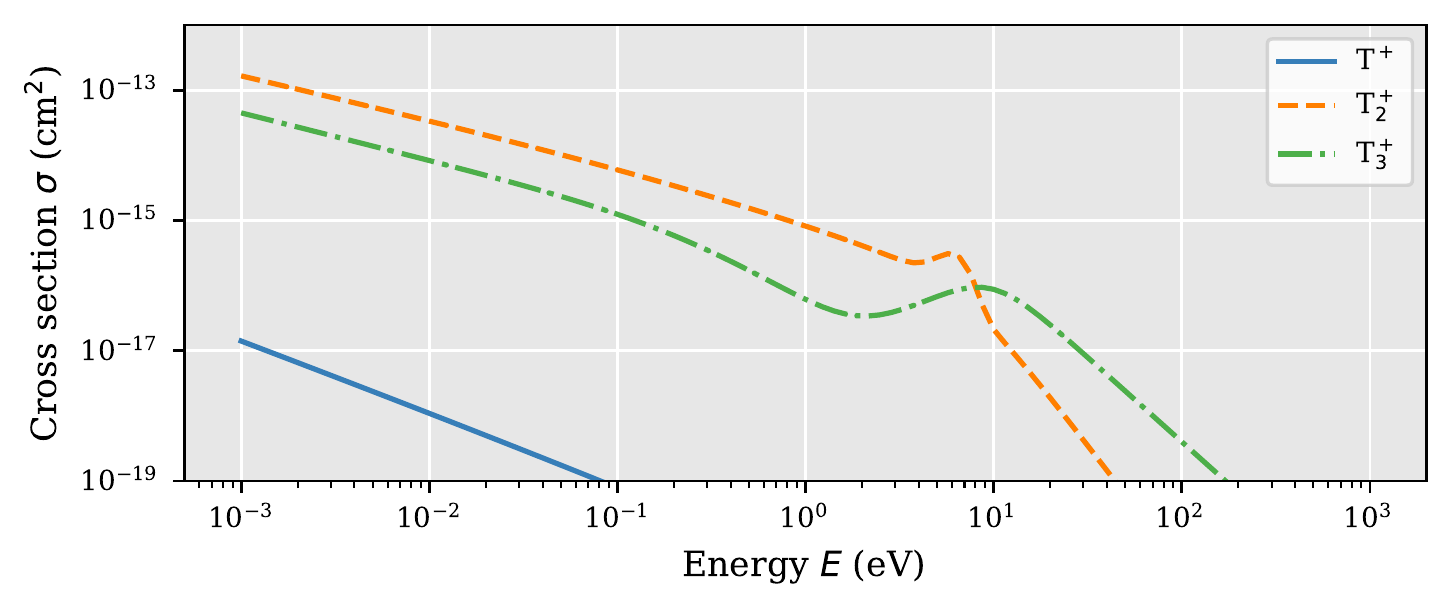}
	\caption{Cross section for electron-ion recombination of different tritium ions. Dissociative recombination cross sections of \ch{T2+} and \ch{T3+} from \citep{Janev:2003aa}. Radiative recombination cross section of \ch{T+} from \citep{Kotelnikov.2019} for comparison. Data provided for the frame of reference, where the ion is at rest.}
	\label{fig:mod:eTI_overview}
\end{figure}

Atomic ions can directly recombine with electrons through radiative recombination, which produces a neutral particle and a photon. The cross section of this interaction is low \citep{Bhattacharyya:1997aa} in comparison to the other recombination channels, see also figure~\ref{fig:mod:eTI_overview}, and it is therefore not considered further in the code.

Tritium molecule ions can undergo two-body recombination and three-body recombination with additional electrons. Matveyev et al. \citep{Matveyev:1995aa} derived a mean rate coefficient
\begin{equation}
	k_\mathrm{3,rec} = \SI{4.7e-31}{\meter \tothe{6} \per \second} \, ,
\end{equation}
for three-body recombination at \SI{80}{\kelvin}. The mean free path of the reaction, assuming even a thousand times higher electron density of $n \approx \SI{1E9}{cm^{-3}}$, is calculated to be on the order of $\mathcal{O} (\SI{{}e3}{\meter})$, which is significantly larger than the length of the source ($L_\text{WGTS} = \SI{16}{m}$ \citep{Aker.352021}). Therefore, three-body recombination is not considered further.

Janev et al. \citep{Janev:2003aa} show that most of the tritium ions undergo dissociative recombination through
\begin{eqnarray}
	\ch{e-} + \ch{T2+} &\rightarrow \ch{T} + \ch{T} \, , \\
	\ch{e-} + \ch{T3+} & \rightarrow \ch{T2} + \ch{T} \, .
\end{eqnarray}
In the code, the cross section of recombination is implemented directly from the analytic formulas of Janev et al. \citep{Janev:2003aa}. The cross section data is depicted in figure~\ref{fig:mod:eTI_overview}.

In the code, the recombination process is implemented as a deletion of the impact particle. The reduction of the density of the corresponding interaction partner is handled by the increase of the additional term $N_\gamma$ in equation~\eqref{eq:target_field_density}. Neutral particles are not tracked in the simulation, and are therefore not considered further.

\subsection{Interactions of tritium ions}
\label{sec:mod:TIT2}

The elastic scattering probability of ions with neutral gas is much larger than for electrons \citep{Janev:2003aa}. Thus, they scatter more often and will most likely adopt the velocity distribution of the neutral gas. Therefore, the energy distribution of ions is not tracked in the simulation, but approximated as a Maxwell-Boltzmann distribution. Only the absolute density is recorded. In the same manner, only ion-gas interactions are implemented, which are relevant at thermal energies.

In the code, three different cluster sizes are distinguished: \ch{T+}, \ch{T2+} and \ch{T3+}. Higher order clusters are neglected here, due to missing cross section data or reaction rates. Similar to \ch{e-}-\ch{T2} interactions, there is no sufficient data on interactions of tritium ions. Models of \ch{H} ions are adopted throughout this section.

\subsubsection*{Elastic Scattering with \ch{T2}}

Tritium ions \ch{T+} and \ch{T3+} can scatter elastically from the neutral gas. At low energies, elastic scattering is the dominant interaction channel. Thus, it is assumed that the ions in the source reach thermal equilibrium quickly. In the code, the elastic cross sections from \citep{Tabata:2000aa} are implemented as analytical formulas.

\ch{T2+} does not scatter elastically with the neutral gas. Nevertheless, these ions perform charge exchange processes with the neutral gas. This way, the \ch{T2+} also adopts the velocity distribution of the neutral gas. In the code, the cross section of charge exchange from \citep{Janev:2003aa} was implemented as an analytical formula.

The kinematic of elastic scattering is treated similar to the one of scattering of electrons with the neutral gas. Also, the guiding center position of the ions is changed in the interaction. The scattering angle is assumed to be isotropically distributed due to the relatively small collision energy.

In the charge exchange process, a new ion is created with a velocity sampled from the velocity distribution of the neutral gas. The position of the initial ion is adopted for the new ion, with the additional shift of the guiding center. The direction of motion is picked at random.

\subsubsection*{Ion cluster formation}

Data from \citep{Gerlich:1990aa} shows that tritium cluster ions exist at \SI{80}{K}. The formation of these clusters is also represented in the code. As already mentioned, three different cluster sizes are distinguished in the code: \ch{T+}, \ch{T2+} and \ch{T3+}.

\ch{T+} can perform radiative clustering or ternary association clustering \citep{Plasil:2012ab}. In the radiative clustering process, the binding energy is released by emitting a photon. In the ternary association clustering process, the additional energy is transferred to a second tritium molecule. There was no data found on the cross section of the clustering processes. Thus, data from the rate coefficient from  \citep{Plasil:2012ab} and \citep{Gerlich:1990aa} was used to calculate a rough estimate of the cross section through equation~\eqref{eq:rate_to_cross_section}.

\ch{T2+} can form \ch{T3+} through collsion with \ch{T2} either by transfering one atom of the \ch{T2} molecule to the \ch{T2+} ion or by transfering a \ch{T+} ion from the \ch{T2+} ion towards the \ch{T2} molecule (directly adapted from proton and atom transfer of hydrogen interactions) \citep{Janev:2003aa}. The sum of the cross sections of both processes is provided by \citep{Janev:2003aa} and implemented as an analytical formula in the code.

In the code, the new ion is created at the same position as the original ion. There was no data available on the differential cross section. Therefore, as a first approximation, the new ions were created with the same direction of motion and the same kinetic energy as the initial ion.

\section{Validation}

The code was tested thoroughly, including a comparison of specific simulations with expectations. These expectations were formulated for scenarios where an analytical expression of the simulation output can be derived. For these simulations, each interaction was investigated detached from the others. The corresponding tests are summarized in the following.

\paragraph{Elastic scattering} Elastic scattering was tested through an injection of monoenergetic particles into a neutral gas reservoir with a Maxwell-Boltzmann distributed velocity. If elastic scattering is implemented correctly, then these particles will adopt the velocity distribution of the neutral gas independent of the initial energy. The tests show that the particles indeed adopt the velocity distribution of the neutral gas. This result was found to be independent on the initial energy of the particles.

\paragraph{Ionization} Ionization was tested through an injection of mononenergetic electrons. The energy of the electrons was chosen in such a way that only one ionization could occur due to the energy loss in the ionization process. The energy distribution of the ionization products were compared to the expected energy distribution. An agreement was found within the limits of the statistical uncertainty. The ionization process can occur either through a dissociative or non-dissociative process. Thus, the number of \ch{T+} ions in relation to the \ch{T2+} ions should match the relation between the cross sections of the reactions. The numbers of the particles were compared and found to be in agreement with the expectations.

\paragraph{Excitation} Excitation was tested through an injection of monoenergetic electrons. In the excitation processes the electrons lose energy depending on the excitation state while keeping all other properties, see section~\ref{sec:mod:electronic_excitation} and \ref{sec:mod:rotation}. Therefore, a comparison of the initial and final energy will result in the energy loss through excitation.

\paragraph{Recombination} was tested through an injection of monoenergetic ions and electrons at one end of the simulation with the energies $E_e$ and $E_i$ respectively. All other interactions were disabled and a magnetic field in the $z$-direction was applied. Thus, there is no dependence on the x and y position. The expected density of the ions $n_i(z, t)$ and electrons $n_e(z, t)$ can be calculated from the coupled continuity equation of electrons and ions with a particle sink. The differential equations in this case then read as
\begin{eqnarray}
	\frac{\partial n_e}{\partial t} &= v_e \frac{\partial n_e}{\partial z} + \sigma(E_e)  n_e  n_i v_e \, ,\\
	\frac{\partial n_i}{\partial t} &= v_i \frac{\partial n_i}{\partial z} + \sigma(E_i)  n_i  n_e v_i \, .
\end{eqnarray}
In the limit of large times it is expected that the density reaches equilibrium. So this coupled differential equation can be solved by the linear equation
\begin{equation}
	n_e(z) = \frac{\sigma(E_e)}{\sigma(E_i)} n_i(z) + c \,
\end{equation}
where $c$ is an integration constant. A linear relation between the electron and ion density is expected. The slope of the test simulation can now be compared to the expectations from the theoretical cross sections. It was found that both values are in good agreement. Additionally, it was recorded how many electrons and ions recombine in the simulation. In the case where electrons and ions have the same velocity, they induce the same density in a simulation without recombination. Thus, in a simulation with recombination it is expected that the number of simulated electrons, which recombine is similar to the number of the simulated ions, which recombine. In a test simulation, it could be shown that the two numbers agree well. In total, it could be shown that the treatment of electron-ion interaction through density fields works as expected.

\section{Physical environment}

\paragraph{Neutral gas density}
The density distribution inside the source is generated by the injection of neutral gas in the center and subsequent pumping of the gas at the pump ports. Thus, the density distribution is shaped by the underlying geometry of the setup, the injection pressure and the pumping rate. A corresponding model was developed by Kuckert et al. \citep{Kuckert.2018}.

\begin{figure}[tb]
	\centering
	\pgfimage{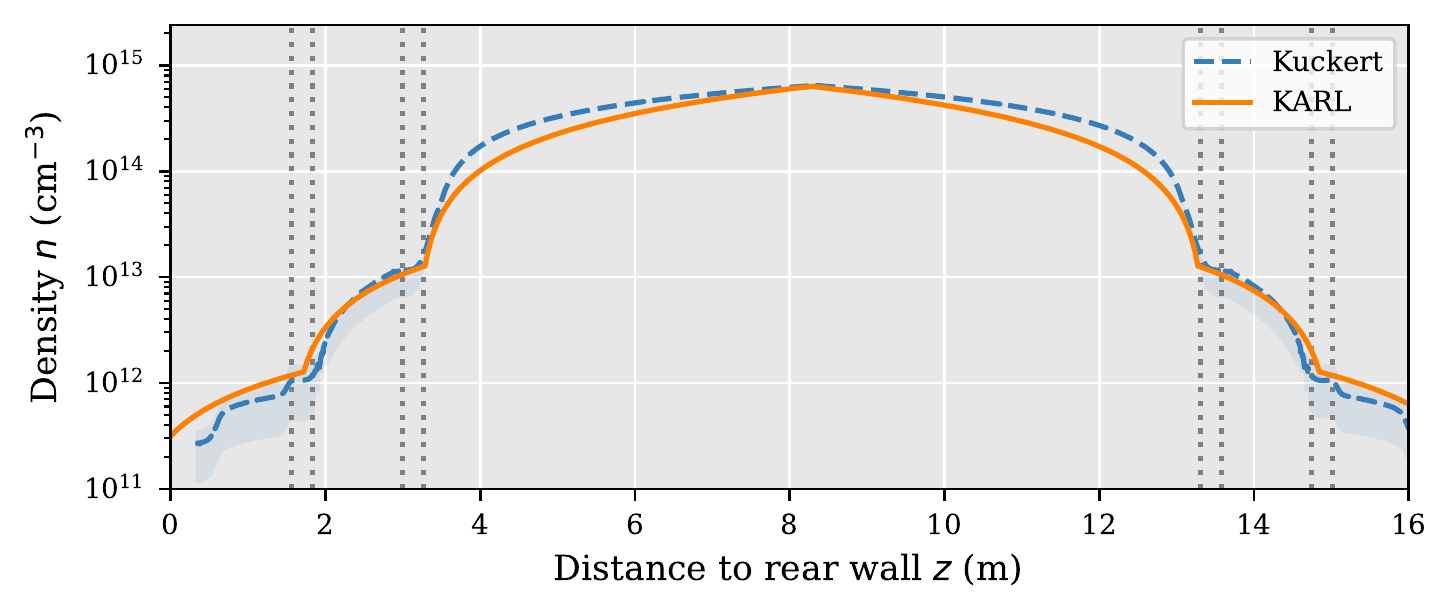}
	\caption{Neutral tritium density from simulation by Kuckert et al. \citep{Kuckert.2018} (maximum scaled to injection density of $n_\text{in} \approx \SI{6.34E14}{cm^{-3}}$  and approximated values implemented in KARL. Horizontal dotted lines represent pump port positions. Tritium is injected at \SI{8.3}{m}.}
	\label{fig:res:n_T2}
\end{figure}

The calculations of Kuckert et al. were performed at the design conditions of the source ($T=\SI{30}{K}$ and $p_\text{in} = \SI{0.337}{Pa}$ \citep{KATRIN2005}). The current measurements of the KATRIN experiments are now performed at higher temperatures and higher injection pressure ($T=\SI{80}{K}$ and $p_\text{in} = \SI{0.7}{Pa}$). This results in a lower injection density of $n_\text{in} \approx \SI{6.34E14}{cm^{-3}}$, using the ideal gas law. It is assumed that the density profile does not change significantly, and the density can be scaled accordingly to the inlet pressure. The resulting density distribution can be found in figure~\ref{fig:res:n_T2}. The calculation of Kuckert et al. were only performed in the direction of the detector. Due to the symmetry of the setup, it can be assumed that the density is distributed similarly in the direction of the rear wall.

A simplified density profile was used for the Monte Carlo simulation. Here, the density is described by six linear functions, see figure~\ref{fig:res:n_T2}.

\paragraph{Neutral gas velocity}
The velocity distribution of the neutral particles is influenced by the inlet pressure and temperature, the pumping speed, and the temperature of the beam tube walls. Each contribution is weighted differently according to the density of the neutral particles. In the center, for example, there are many molecule-molecule interactions. Here, the velocity distribution is dominated by the particle flow. In the rear wall chamber, the density is reduced drastically, such that molecule-molecule interactions become negligible and collisions with the tube walls are dominant.

In the viscous flow regime, the velocity distribution can be described by a Maxwell-Boltzmann distribution with an additional bulk velocity in the longitudinal direction \citep{Kuckert.2016}. The temperature of this distribution is given by the temperature of the beam tube walls. The bulk velocity describes the streaming process from the inlet in the center of the WGTS to both sides. Simulations of Kuckert et al. \citep{Kuckert.2018} show that the bulk velocity is dependent on the longitudinal as well as the radial position. The bulk velocity increases from the inlet to both sides. Radially, the bulk velocity is reduced at the tube walls, corresponding to wall interactions which slow down the gas movement.

At lower densities, the velocity can no longer be described through a drifting Boltzmann distribution, because interparticle collisions are reduced. In first order, the particles follow a Maxwell-Boltzmann distribution, with a temperature of the beam tube walls.

\paragraph{Magnetic field}
The magnetic field inside the source is generated by nine super conducting magnets. These magnets surround each tube element of the source. These magnets provide a strong field and are set to $\SI{2.5}{T}$ with a low drift $<\SI{0.03}{\percent}$ per month \citep{Arenz.2018b} and fluctuations below \SI{0.2}{\percent} per data acquisition run \citep{Aker.352021}. For the simulation, the magnetic field is assumed to be homogeneous throughout the source.

\section{Results}
\label{sec:results}
The results presented below were generated through a simulation with a maximal tritium density of \SI{6.3E14}{cm^{-3}} at a temperature of \SI{80}{K}. The source tube was approximated by a cylinder with a length of \SI{16}{m} and a radius of \SI{4.5}{cm}, simplifying the segmentation of the source, see section~\ref{sec:mon:density_fields}. The injection of tritium gas is assumed to be at the center of the source tube; \SI{8}{m} from the rear wall. No electric background field was used, and the magnetic field was set to \SI{2.5}{T}. \num{E6} $\upbeta$-decays were simulated, resulting in a statistical uncertainty of \SI{0.1}{\percent}, see section~\ref{ssec:particle_tracking_and_diagnostics}. This uncertainty will be used for all simulation results presented below. The source was segmented into 32 z bins and 15 radial bins. No segmentation was used in the azimuthal direction due to the expected azimuthal symmetry of the result. Density and currents were recorded at 100 energy bins with a logarithmic subdivision.

\begin{figure}[tb]
	\centering
	\pgfimage{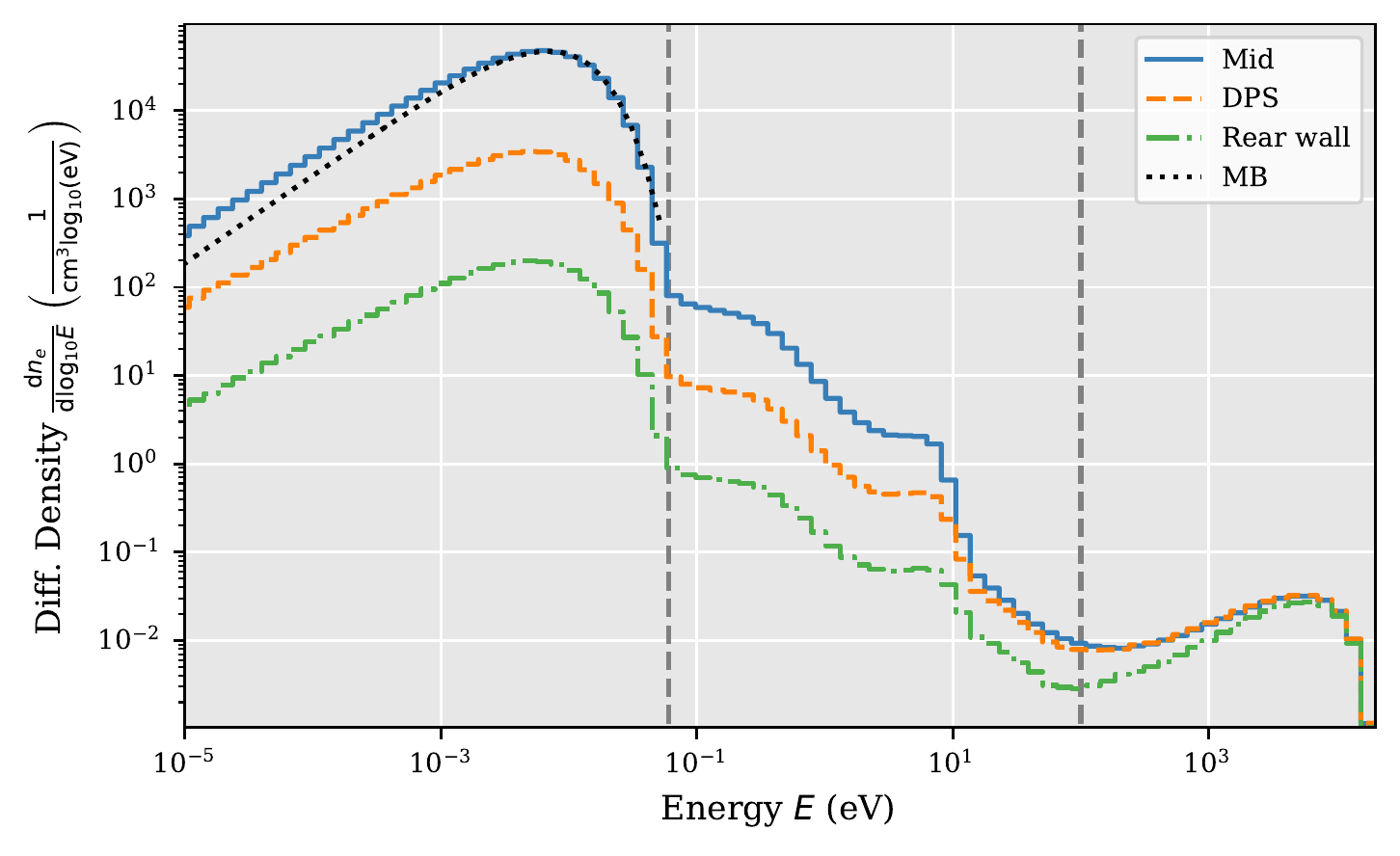}
	\caption{Electron spectrum of three different segments in the source. Mid: central segment connected to the inlet, Rear wall: segment in front of the rear wall, DPS: segment next to the DPS. MB: Normalized Maxwell-Boltzmann distribution. Vertical lines for distinction between energy regions.}
	\label{fig:res:n_e_E}
\end{figure}

\paragraph{Electron spectrum}
The electron spectrum was recorded at \num{32} different positions in the source. Three distinct positions are compared here: a segment, which is directly connected to the tritium inlet; a segment, which is next to the rear wall; and a segment, which is connected to the DPS. The corresponding spectra can be found in figure~\ref{fig:res:n_e_E}. Three different energy regions can be identified (distinguished by vertical lines in the figure): a so-called thermal region ($E < \SI{6E-2}{eV}$), where the electrons reached thermal equilibrium with the neutral gas, a so called $\upbeta$-decay region ($E > \SI{E2}{eV}$), where the spectrum is governed by the Fermi distribution of $\upbeta$-decay and a so-called secondary region in between the other two regions.

The spectrum in the $\upbeta$-decay region shows little deviation between the three different positions. Therefore, the $\upbeta$-electrons either travel through the source unobstructed, or lose the main portion of their energy through ionization and subsequent scattering. There is a slight difference at the lower end of the $\upbeta$-region between the rear wall and the other two positions. Electrons which reach the rear wall segment can have traveled at maximum two times through the source through reflection at the spectrometer, while electrons at the DPS traversed the source only once. Therefore, more $\upbeta$-electrons have scattered before reaching the rear wall segment, which reduces the density in this energy range.

The spectral shape in the thermal region is similar for each segment. These electrons have scattered multiple times with the neutral gas and thus adopted the temperature of the gas. The shape can be described by the Maxwell-Boltzmann distribution \citep{peckham.1992}, also depicted in figure~\ref{fig:res:n_e_E}, scaled by the factor $\sqrt{E}$ taking into account the logarithmic binning and the speed of particles leaving the simulation domain. Differences at the low energy side arise from the reduced cross section of elastic scattering at lower energies, see figure~\ref{fig:mod:eT2_overview}. Particles with lower energies scatter less, increasing their time spent in the simulation domain, which in turn manifests in a higher density, compare equation~\eqref{eq:target_field_density}. This behavior was verified in targeted test simulations with fixed cross section, which resulted in a good agreement of the Maxwell-Boltzmann distribution with the simulated data in the lower energy regime down to \SI{E-6}{eV}.

The similarity to the Maxwell-Boltzmann distribution also shows that recombination is only a secondary effect. Otherwise, there would be much fewer particles at low energies, where the cross section of recombination is increased significantly, see figure~\ref{fig:mod:eTI_overview}.

The absolute density is the largest in the center, while it decreases towards both sides. In the center, there are many collisions. Thus, the electron density is increased here. The density at the DPS is higher than at the rear wall. This is caused by the reflection of electrons at the DPS, see section~\ref{ch:katrin}.

The secondary region shows two distinct features. A continuous decrease from \SI{5E-2}{eV} to \SI{10}{eV} and an edge at \SI{10}{eV}. The energy of the edge is consistent with the threshold energy for ionization and electronic excitation. Electrons with a lower energy can no longer take part in these interactions. The continuous decrease arises as a combination between elastic scattering and rotational and vibrational excitation. The rotational excitation has a threshold energy of approximately \SI{5E-2}{eV}, which is consistent with the end of the secondary region, see figure~\ref{fig:mod:eT2_overview}.

\begin{figure}[tb]
	\centering
	\pgfimage{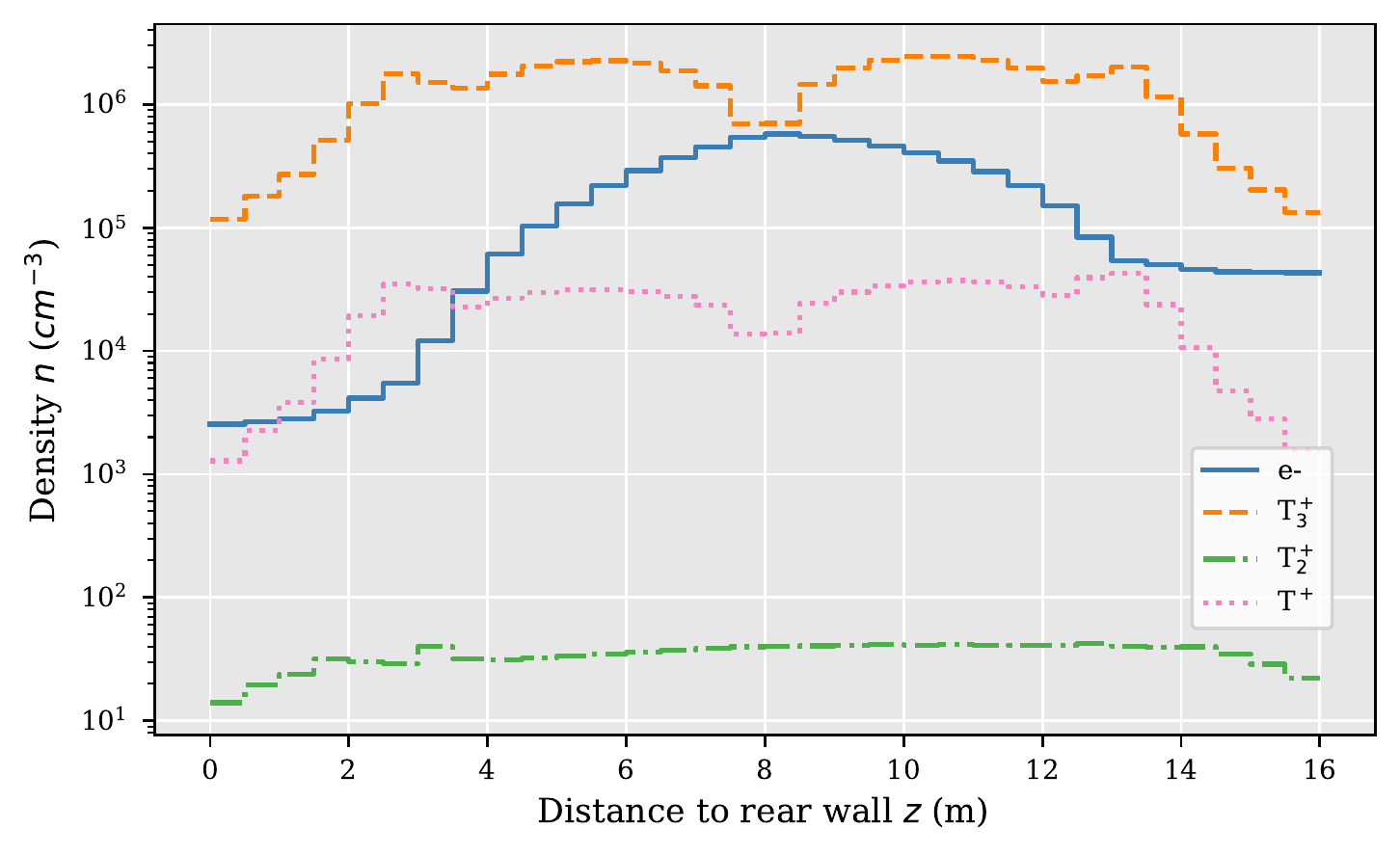}
	\caption{Particle densities in longitudinal direction. In the simplified geometry, Tritium is injected at a point \SI{8}{m} away from the rear wall.}
	\label{fig:res:n_long_all}
\end{figure}

\paragraph{Particle densities}
The density of each particle species can be derived from the sum over the density spectrum. The corresponding distribution along the source tube can be found in figure~\ref{fig:res:n_long_all}. It can be seen that the density profile of each specie is different from each other. The absolute charge density is non-zero in this simulation, despite the fact that electrons and ions are created in pairs. This is not surprising because electrons scatter less and are moving faster, and therefore leave the source tube faster than the ions. Therefore, their density is reduced in comparison to the ions. In the experiment, it can be assumed to first order that there exists quasi neutrality in the source. Thus, the density differences between electrons and ions will induce an electric field. This field will be calculated in a different dedicated simulation.

The electron density is dominated by the thermal part of the spectrum. $\upbeta$ and secondary electrons make up only a maximum of \SI{4E2}{cm^{-3}}. The electron distribution shows a maximum in the center of the source and falls off to both sides. The probability of scattering is the highest in the center of the source. Hence, the electron density is the largest here. The scattering probability is decreased to both sides as the neutral gas density decreases. The electron density is asymmetric towards the inlet, despite the symmetric shape of the density. This is caused by the reflection of electrons at the DPS, which increases the density in this region. Nevertheless, the density before the DPS is not increased drastically, which suggests that electrons can pass the high density region of the inlet.

The ion distribution is different for the three different ion species. The most abundant ions are \ch{T3+} ions. This is expected due to the high conversion cross section of \ch{T+} and \ch{T2+} towards \ch{T3+}. The density of the \ch{T3+} ions is governed by ionization through electron impact and by elastic scattering with the neutral gas. Ionization is most probable in the center, where the neutral gas density is the highest. The resulting density of ionization is superposed by the streaming process, caused by the elastic scattering. The motion of ions is aligned with the bulk velocity of the gas. Thus, the ions are transported to both sides of the source, increasing the density in these regions. The density drops in the region in front of the rear wall and in front of the DPS. This is caused by the reduction of the neutral gas density and the drop out of the neutral gas drift.

The density of \ch{T+} is distributed similar to the density of \ch{T3+}. The shape of the spectrum is created by the same effects as for \ch{T3+}, namely elastic scattering with neutral gas and particle transport. The density of \ch{T+} is lower than the density of \ch{T3+}. In this simulation, \ch{T+} can only be created through dissociative ionization. The cross section of this reaction is significantly lower than for non-dissociative ionization. Therefore, less \ch{T+} is created in comparison to \ch{T3+}, which is created through the efficient conversion of \ch{T2+}. Additionally, \ch{T+} can be converted to \ch{T3+} through ternary association, which lowers the density of \ch{T+} in comparison to the density of \ch{T3+}.

The density of \ch{T2+} is significantly different from the other two ion species. The shape is almost constant, and the density is significantly lower. This behavior can be explained by the combination of two effects. First, the cross section of cluster formation is higher than the cross section for charge transfer \citep{Janev:2003aa}. Therefore, \ch{T2+} ions exist only for a short time between creation and their first interaction. Thus, the density of \ch{T2+} is significantly lower. Secondly, the probability for cluster formation is dependent on the tritium density, while the creation mechanism through beta decay and ionization is also dependent on the tritium density. In total, both effects cancel out and the \ch{T2+} density therefore shows no large longitudinal variation. Similar to \ch{T3+} and \ch{T+}, there is a slight decrease of the density in front of the rear wall and the DPS. Again, the mean free path of the particles has increased enough that some particles can leave the source, which reduces the density.

\paragraph{Particle currents}

\begin{figure}[tb]
	\centering
	\pgfimage{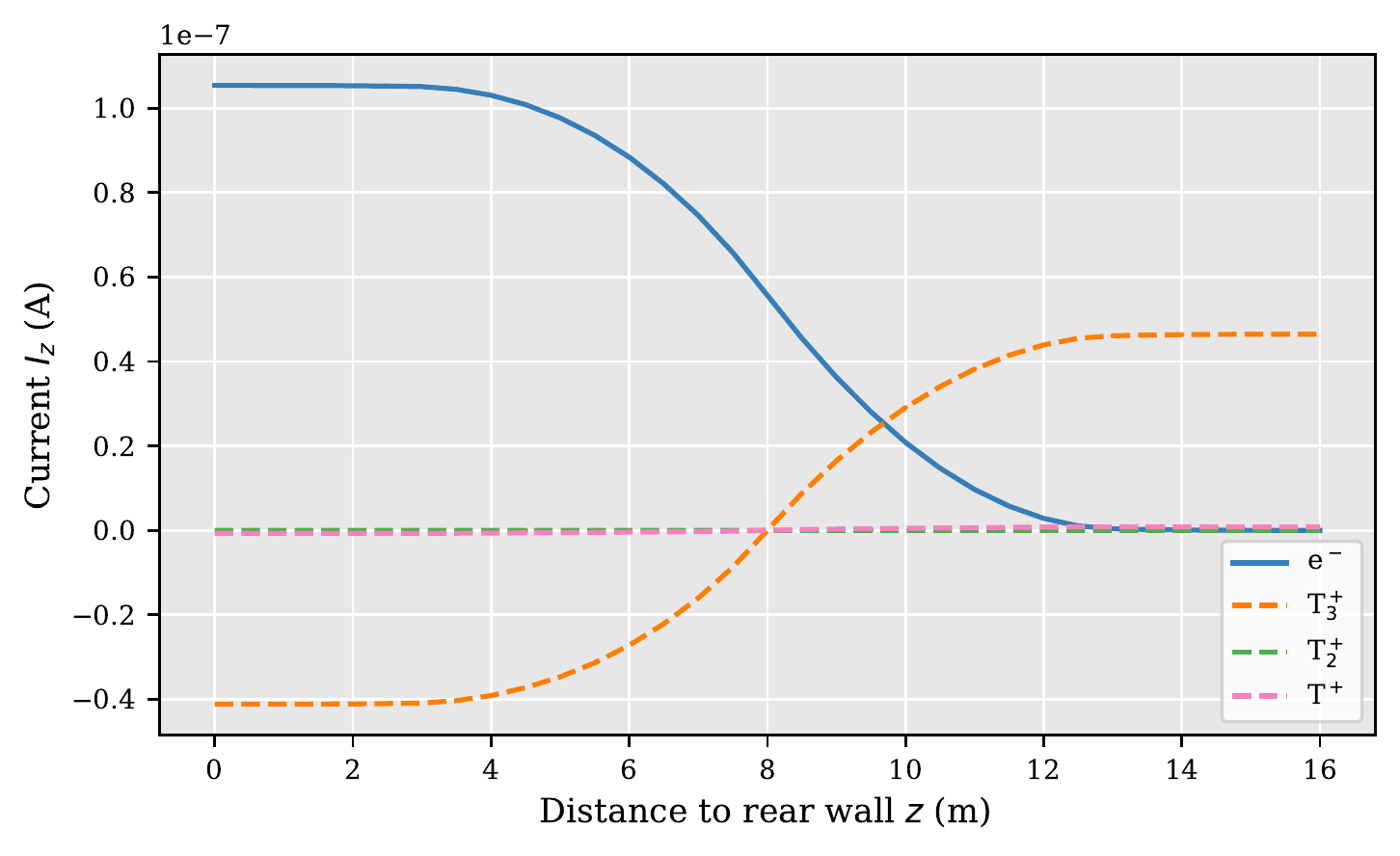}
	\caption{Particle current in longitudinal direction. The sign of the current corresponds to the direction of the particle flow multiplied with the sign of the particle charge. In the simplified geometry, Tritium is injected at a point \SI{8}{m} away from the rear wall.}
	\label{fig:res:Iz_all}
\end{figure}

The particle current was determined through \num{33} virtual planes evenly spaced longitudinally. The planes were aligned in such a way that the first and last longitudinal plane coincide with the rear wall and the passage to the DPS. The corresponding longitudinal current profile can be found in figure~\ref{fig:res:Iz_all}.

The shape of the electron current coincides with expectations. Electrons are reflected at the DPS by the potential of the dipole electrodes and the spectrometer, see section~\ref{ch:katrin}. The corresponding current is therefore zero. Electrons are most likely to be terminated at the rear wall. Thus, the electron current is the highest here. The electron current decreases continuously from the rear wall towards the DPS. Hence, the electrons can stream from the upstream side towards the downstream side and are not hindered from passing by the neutral gas flow.

The ions are created in the center and stream to both sides. There is no driving force of the ions towards the other side of the inlet. Thus, the ion current at the inlet (\SI{8}{m} in the simplified geometry) is zero. The current leaving towards the DPS and towards the rear wall is almost identical. The differences can be explained by the increase of $\upbeta$-electrons through reflection at the DPS side.

The current of both electrons and ions reaches a plateau between \SIrange{0}{2}{m}, and \SIrange{14}{16}{m}. This behavior is directly linked to the reduced neutral gas density in these regions, see figure~\ref{fig:res:n_T2}. A negligible amount of ions and electrons are created here through beta decay and ionization. Thus, the current in these regions is a result of the creation of electrons and ions in the center of the source and the subsequent magnetic guiding towards the plateau regions. Hence, the current does not increase further in these regions.

\begin{figure}[tb]
	\centering
	\pgfimage{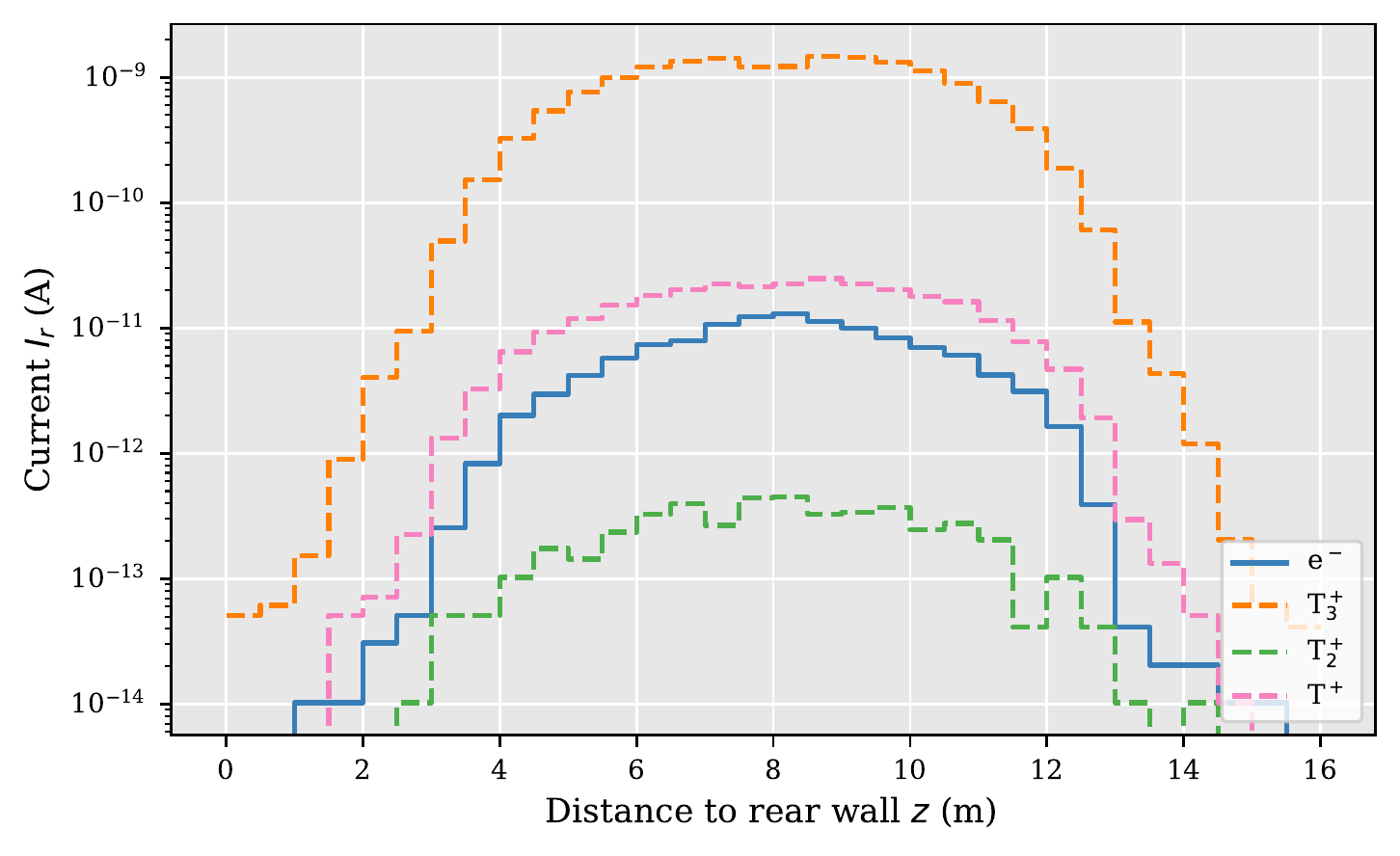}
	\caption{Absolute value of the radial particle current towards the beam tube as a function of longitudinal position along the beam tube. The plateaus mark the longitudinal bin width, where the total current flows through. In the simplified geometry, Tritium is injected at a point \SI{8}{m} away from the rear wall.}
	\label{fig:res:j_tube_all}
\end{figure}

The sum of the ion current leaving the source at the DPS and at the rear wall is below the electron current leaving at the rear wall. This can only be explained through an ion current, which is directed towards the beam tube. The corresponding current, displayed in figure~\ref{fig:res:j_tube_all}, was detected in the simulation through radial virtual planes located at the beam tube walls. This current is caused by the position change of the guiding center after elastic scattering. The position change is dependent on the Larmor radius of the particle, see equation~\eqref{eq:position_change_el_scattering}, which in turn scales with the square root of the mass. Additionally, the cross section of ion scattering is much higher than for electron scattering. Therefore, more collisions occur, which can shift the guiding center, increasing the maximal position change in radial direction \citep{Chen.1984}. Because of the larger position change and larger scattering probability, the radial ion current is expected to be significantly higher than the radial electron current, which is also reflected in the simulation results.

The absolute current to the tube is also influenced by the density profile of electrons and ions. It can be seen that the shape of the current towards the beam tube is similar to the shape of the neutral particle density. This is caused by the increased elastic scattering probability in the center due to the increased neutral gas density. The \ch{T+} current is below the \ch{T3+} current, which is caused by the density difference and the different Larmor radius of the two ion species. The current of \ch{T+} is of the same order of magnitude as that for the electrons, even though the masses of both species is significantly different. This behavior is caused by the significant difference in the density.

There is a contribution of \ch{T2+} to the radial current. This current stems manly from \ch{T2+} ions, which are created close to the beam tube walls. Otherwise, they would have transformed into the other ion species through cluster formation processes, because the cross section of cluster formation is higher than the cross section for charge transfer \citep{Janev:2003aa}.

\section{Discussion}

\begin{figure}[tb]
	\centering
	\pgfimage{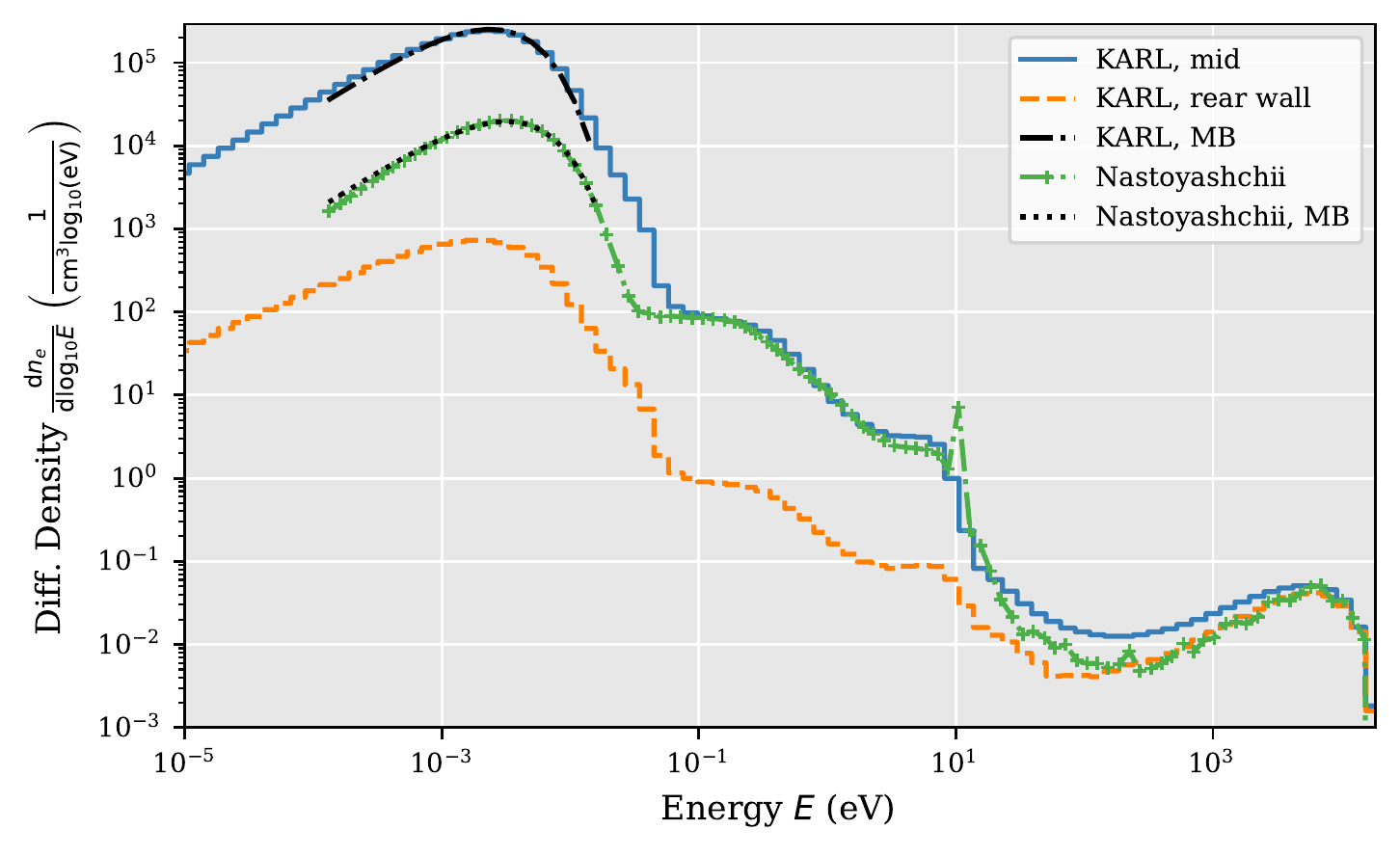}
	\caption{Comparison of the position independent result of Nastoyashchii et al. \citep{Nastoyashchii:2005aa} and the position dependent results of this paper (mid: central segment connected to the inlet, rear wall: segment in front of the rear wall). The simulation was performed at the same settings ($T = \SI{30}{K}$ and $n_{T_2} = \SI{1E15}{cm^{-3}}$). No explicit comment was made on the magnetic field strength by Nastoyashchii et al. Here we assumed it to be $B=\SI{3.6}{T}$. The results of Nastoyashchii et al. were scaled such that the density at the high energy region coincides with our simulation. MB: Normalized Maxwell-Boltzmann distribution.}
	\label{fig:dis:comparison_nastoyashchii}
\end{figure}

The latest results of the electron spectrum in the KATRIN source was published by Nastoyashchii et al. \citep{Nastoyashchii:2005aa} in 2005. They derived an electron spectrum of the KATRIN source through a Monte Carlo simulation. The spectrum was derived through an evaluation of the total time particles spent in the entirety of the source. Position dependence of the spectrum was not included, and therefore the spectrum can only be construed in first order as a mean spectrum of the electrons in the source. In this paper, we derived electron spectra at different positions in the source, see figure~\ref{fig:res:n_e_E}. A comparison of our results to Nastoyashchii et al. is depicted in figure~\ref{fig:dis:comparison_nastoyashchii}. Nastoyashchii et al. only derived a relative spectrum and did not specify the absolute scale. Here, we assume, that the high energy region of their and our simulation coincide. Thus, their spectrum is scaled accordingly, such that the data point of Nastoyashchii et al. at \SI{6886}{eV} has the same value as the energy bin from \SIrange{5447}{7060}{eV} of our calculation from the central position. It can be seen that the distribution of Nastoyashchii et al. shows similar features as our simulation: a thermal region, a $\upbeta$-region and a secondary region. However, there exist significant differences, which will be discussed in the following.

The results of Nastoyashchii et al. only provide a mean particle density of the source. It can be seen in our simulations that there exist significant differences in the spectrum at the different positions in the source. These differences might change the description of the plasma. Nastoyashchii et al., for example, assume implicitly in their affiliated calculation of the plasma potential that the high energy electrons of the spectrum can be neglected and it can solely be described by a thermal distribution (equation~(1) in their publication). However, our simulations show that this assumption is not true for every position in the source, especially when considering that the maximum energy density in the beta region ($\epsilon \sim \SI{10}{eV cm^{-3}}$) is higher than the energy density of the thermal region in front of the rear wall ($\epsilon \sim \num{1}$ $\text{eV cm}^{-3}$). Thus, a different plasma description is necessary, which must also include high energy electrons.

There exists a spike (one data point) at \SI{10}{eV} in the results of Nastoyashchii et al., which does not exist in our simulation, see figure \ref{fig:dis:comparison_nastoyashchii}. The existence of this peak is important, because there might be plasma instabilities linked to it (Penrose criterion \citep{mouhot.2011}). Nastoyashchii et al. do not specify the origin of this peak and provide no information on the statistics of the simulation. We did not find any physical explanation of the peak. To test whether statistical fluctuations could produce such a peak, we performed our simulations with different random number seeds. No peak was found.

Although the thermal region of Nastoyashchii et al. seems to have the same shape and position as our calculations, there is a difference between both. For comparison, a least squares fit of a Maxwell-Boltzmann distribution \citep{peckham.1992}, scaled by the factor $\sqrt{E}$ taking into account the logarithmic binning and the speed of particles leaving the simulation domain, was performed on both datasets in the energy range of \SIrange{1.3E-4}{2.0E-2}{eV}, resulting in $T_\text{Nast.} = \SI{37.2}{K}$ and $T_\text{Karl} = \SI{27.7}{K}$. The corresponding distributions are also depicted in figure~\ref{fig:dis:comparison_nastoyashchii}. The expected temperature is \SI{30}{K}, the same as the temperature of the tritium gas. The slightly lower temperature and increased density in the lower energy regime of our calculation can be retraced to the energy dependency of the elastic cross section, confirmed by test simulations, compare to section~\ref{sec:results}. The result of Nastoyashchii et al. shows both a higher temperature and lower densities in the lower energy regime than expected. It is unclear, which mechanism could cause this behavior, and Nastoyashchii et al. did not specify, where the discrepancy comes from.

\begin{figure}[tb]
	\centering
	\pgfimage{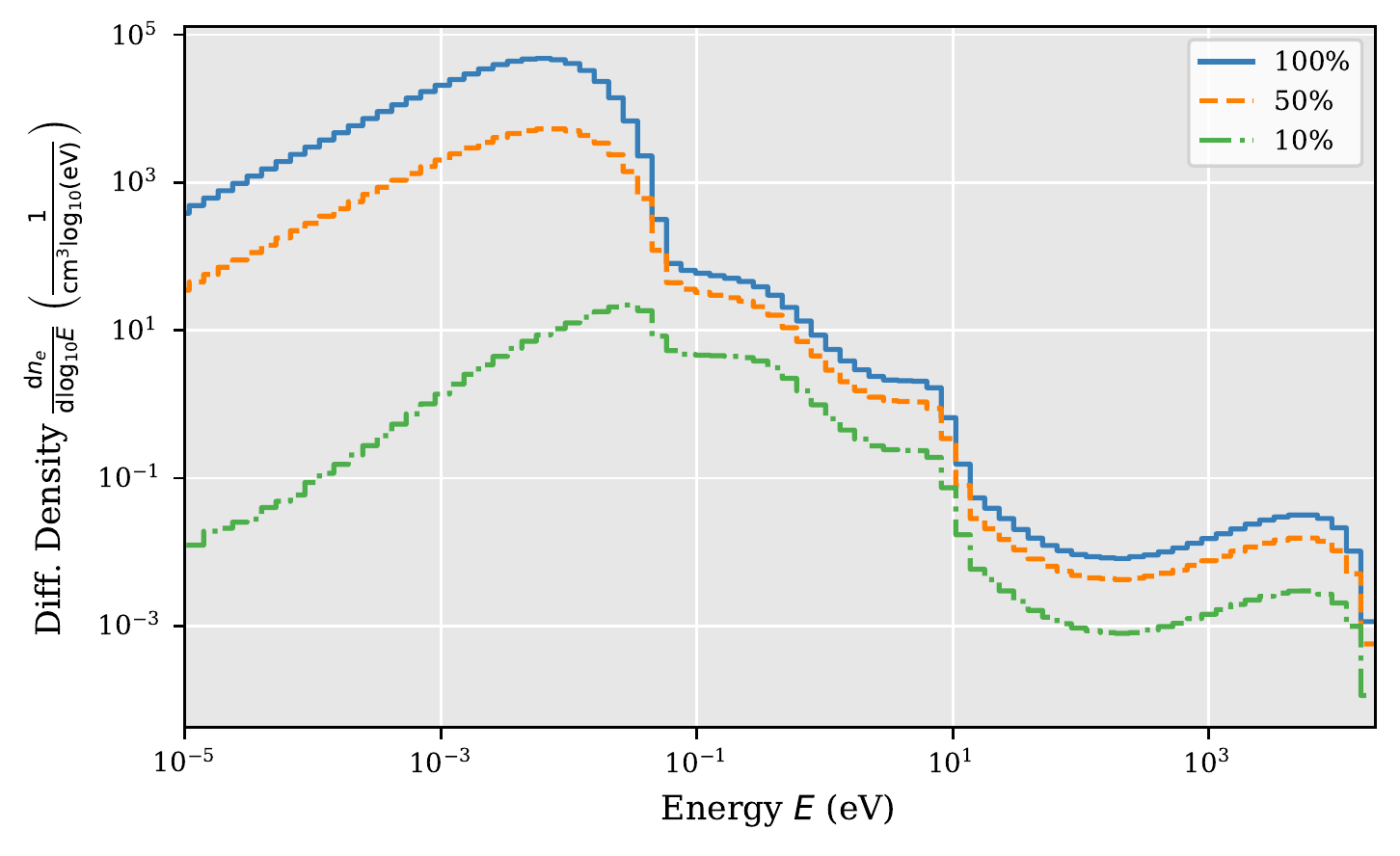}
	\caption{Comparison of the electron spectrum at different inlet densities. \SI{100}{\percent} correspond to an inlet density of \SI{6.3E14}{cm^{-3}}. The source temperature was set to $T=\SI{80}{K}$ and the magnetic field strength to $B=\SI{2.5}{T}$.}
	\label{fig:dis:comparison_densities}
\end{figure}

The simulation of Nastoyashchii et al. were performed at fixed source conditions ($T = \SI{30}{K}$ and $n_{T_2} = \SI{1E15}{cm^{-3}}$). No further comment was made on the magnetic field strength. In this paper, we assumed that the magnetic field strength of Nastoyashchii et al. was set to $\SI{3.6}{T}$, the same value as it was stated in the design report \citep{KATRIN2005}. The measurement conditions have changed since 2005 \citep{Aker.352021} and might still be changed in the future \citep{aker2021direct}. Our new simulation code allows now for an easy adaptation of the simulation to the source parameters used in the measurement. The simulation can also be used to probe favorable source parameters. One example is the investigation of the inlet pressure and therefore inlet density, see figure~\ref{fig:dis:comparison_densities}. It can be seen that the general shape of the spectrum does not change significantly, when reducing the density to half of its original value. However, at \SI{10}{\percent} of the original value, the shape of the spectrum changes significantly. A classification into a distinct thermal and secondary region is no longer possible. Thus, it can be assumed that the plasma physics changes significantly and a comparison of neutrino mass measurement data of such a low density can not be compared without special treatment to measurement data at nominal inlet density.

\begin{figure}[tb]
	\centering
	\pgfimage{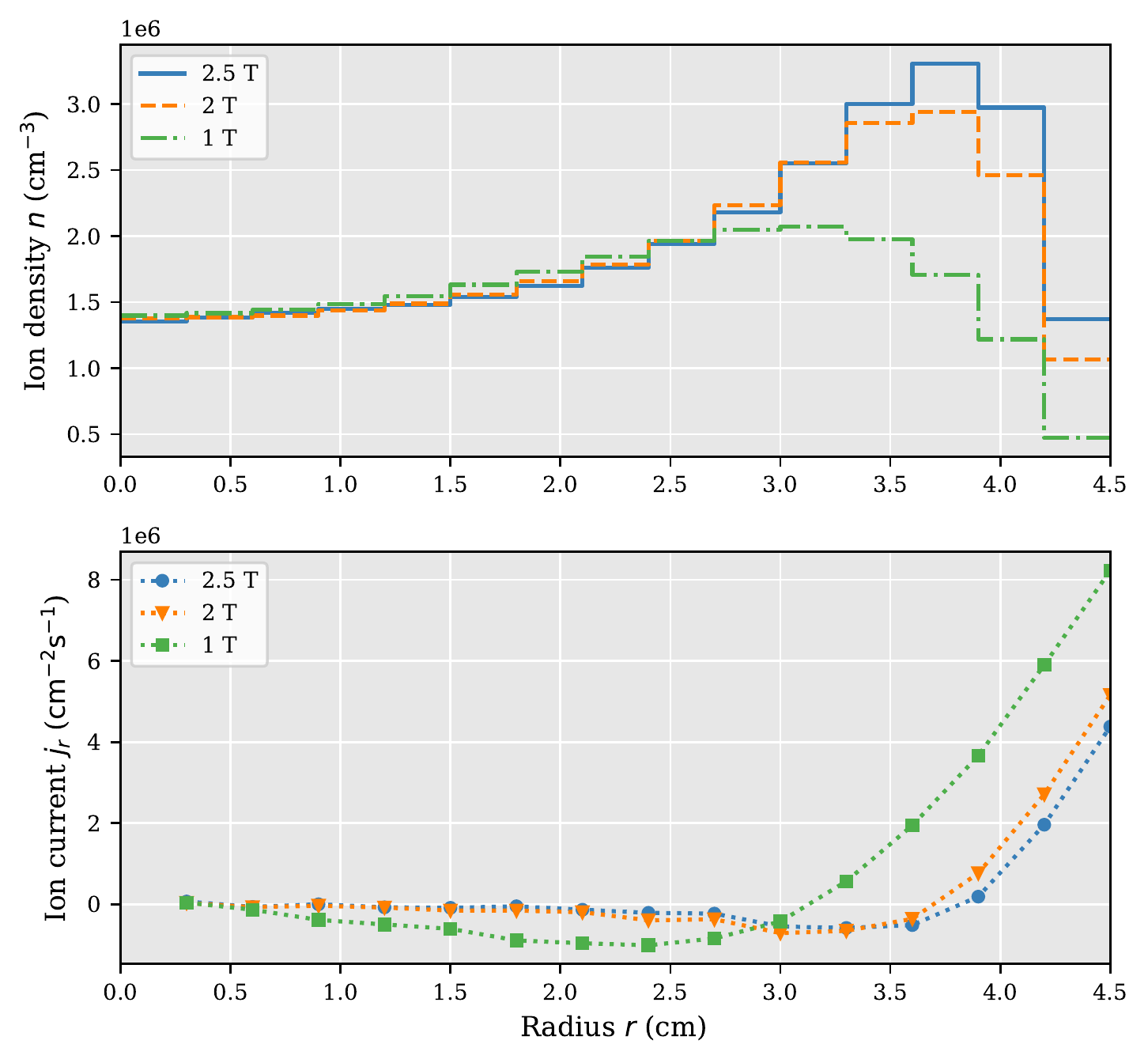}
	\caption{Comparison of the radially resolved ion density and current density at different magnetic field strength at \SI{5}{m} distance to the rear wall. The source temperature was set to $T=\SI{80}{K}$ and the inlet density to \SI{6.3E14}{cm^{-3}}.}
	\label{fig:dis:comparison_magnetic_field}
\end{figure}

One observable of the KATRIN source plasma is the current leaving the source, measured at the rear wall and at the DPS dipole electrodes \citep{Aker.352021}. The sum of these currents is in general not zero, which means that there is a current contribution towards the beam tube \citep{Friedel.2020}. Our radially resolved simulations could now show that there is a non-negligible current contribution towards the beam tube walls, which is caused by the gyro center movement through elastic scattering, see figure~\ref{fig:res:j_tube_all}. This is an improvement over the calculation of Nastoyashchii et al., which did not calculate particle currents in their simulation, especially not in radial direction. Our simulations can also be used to investigate the behavior of this current, especially under a change of the source magnetic field strength, see figure~\ref{fig:dis:comparison_magnetic_field}. It can be seen that the calculated current towards the beam tube is dependent on the source magnetic field. This behavior is expected, because the gyro radius of the particles is increased at lower magnetic field strength, and thus increasing the position change per collision, see equation~\eqref{eq:position_change_el_scattering}.

The simulation with varied magnetic field also shows significant differences in the radial particle densities and particle currents, see figure~\ref{fig:dis:comparison_magnetic_field}. These densities and currents can now be used to investigate the radial profile of the potential in the source through dedicated field solvers. This theoretical investigation can be supplemented by measurements with the radially resolved detector \citep{Aker.352021}. Thus, an experimental investigation of the simulation results will be possible.

\section{Conclusion and outlook}
In this paper, we presented the first results of a new Monte Carlo code, which simulates the density and energy distributions of charged particles in the KATRIN source. The distributions were obtained through tracking single particles in the source until they either recombine or are absorbed by the metal surfaces. The single particles can interact indirectly with each other through density fields, see section~\ref{sec:mon:density_fields}. Therefore, this approach allows for the simulation of electron-ion interactions alongside electron-neutral and ion-neutral interactions.

The simulated electron spectra showed three regions: a $\upbeta$-region, a thermal region and a secondary region. The relative density of these regions was found to be dependent on the position in the source, see figure~\ref{fig:res:n_e_E}. We also determined the total density of electrons and ions depending on the position in the source, see figure~\ref{fig:res:n_long_all}. Significant differences were found between the densities of the different species. Additionally, we determined the particle current depending on the position in the source through virtual barriers, see figures~\ref{fig:res:Iz_all} and \ref{fig:res:j_tube_all}. A non-negligible ion current towards the beam tube was found.

Our simulations are a significant improvement to the previous simulation results of Nastoyashchii et al. \citep{Nastoyashchii:2005aa}. Notably, the electron spectrum can be determined depending on the position in the source. Additionally, we can now investigate the particle behavior at various source conditions, which can be directly compared to measurement data. The source conditions include among others the magnetic field strength, the gas temperature and the neutral particle density. Investigations of the influence of the interactions and their cross section can also be performed.

The simulations presented in this paper were performed without an electric field in the source. This field arises in the experiment from surface potential differences and through the plasma generated by electrons and ions. The results of this paper can provide a starting point of the electric field calculations. The interplay of electric field and particle density is very intricate due to the nature of plasmas and demands special treatment in the future.

\acknowledgments

F.S. would like to thank the Deutsche Forschungsgemeinschaft (DFG) for support through Grant No. SP 1124/9. J.K. acknowledges the support of the Ministry for Education and Research BMBF (5A17PDA, 05A17PM3, 05A20VK3). The authors gratefully acknowledge the Gauss Centre for Supercomputing e.V. (www.gauss-centre.eu) for funding this project by providing computing time on the GCS Supercomputer SuperMUC-NG at Leibniz Supercomputing Centre (www.lrz.de). We are grateful to F. Glück, C. Reiling and G. Drexlin for their support and discussions.

\bibliographystyle{JHEP}
\bibliography{ref}

\end{document}